\documentclass[aps,%
               prb,%
               reprint,%
               showpacs,%
               superscriptaddress,%
               longbibliography,%
               floatfix,%
               nofootinbib%
               ]{revtex4-2}

\usepackage[utf8]{inputenc}
\usepackage[usenames,dvipsnames]{xcolor}
\usepackage{graphicx}
\usepackage{amsfonts}
\usepackage{amssymb}
\usepackage{amsmath}
\usepackage{mathtools}
\usepackage{bm}
\usepackage{dsfont}
\usepackage{braket}

\usepackage[pdfusetitle,bookmarks=true,colorlinks,citecolor=blue,urlcolor=blue]{hyperref}

\newcommand{\cre}[2]{{#1}^{\dagger}_{#2}}
\newcommand{\ann}[2]{{#1}^{\phantom{\dagger}}_{#2}}

\newcommand\bs[1]{\ensuremath{\boldsymbol{#1}}}

\newcommand{\eq}[1]{Eq.\thinspace(\ref{#1})}
\newcommand{\eqs}[2]{Eqs.\thinspace(\ref{#1},\ref{#2})}

\newcommand{\fig}[1]{Fig.\thinspace{}\ref{#1}}
\newcommand{\fc}[1]{({#1})}
\newcommand{\figc}[2]{Fig.\thinspace{}\ref{#1}\thinspace{}\fc{#2}}

\newcommand{\TUM}{\affiliation{Technical University of Munich, TUM School of Natural Sciences, Physics Department, 85748 Garching, Germany}}
\newcommand{\MCQST}{\affiliation{Munich Center for Quantum Science and Technology (MCQST), Schellingstr. 4, 80799 M{\"u}nchen, Germany}}
\newcommand{\Harvard}{\affiliation{Department of Physics, Harvard University, Cambridge, MA 02138, USA}}
\newcommand{\CAL}{\affiliation{Department of Physics, University of California, Berkeley, CA 94720, USA}}

\begin{document}

\title{Fractonic Luttinger Liquids and Supersolids in a Constrained Bose-Hubbard Model}
\author{Philip Zechmann}
\TUM \MCQST
\author{Ehud Altman}
\CAL
\author{Michael Knap}
\TUM \MCQST
\author{Johannes Feldmeier}
\TUM \MCQST \Harvard

\date{\today}

\begin{abstract}
Quantum many-body systems with fracton constraints are widely conjectured to exhibit unconventional low-energy phases of matter. In this paper, we demonstrate the existence of a variety of such exotic quantum phases in the ground states of a dipole-moment conserving Bose-Hubbard model in one dimension. 
For integer boson fillings, we perform a mapping of the system to a model of microscopic local dipoles, which are composites of fractons. We apply a combination of low-energy field theory and large-scale tensor network simulations to demonstrate the emergence of a dipole Luttinger liquid phase. At non-integer fillings our numerical approach shows an intriguing compressible state described by a quantum Lifshitz model in which charge density-wave order coexists with dipole long-range order and superfluidity -- a ``dipole supersolid''. While this supersolid state may eventually be unstable against lattice effects in the thermodynamic limit, its numerical robustness is remarkable. We discuss potential experimental implications of our results.
\end{abstract}

\maketitle
   
\section{Introduction} \label{sec:introduction}
The current advent of quantum simulation technology is marked by rapid progress in controlling strongly interacting many-body systems. 
In particular, the ability to engineer highly specific quantum Hamiltonians has raised immense interest in the physics of quantum systems subjected to dynamical constraints.
A particularly exciting class of systems that has caught much attention in this regard are so-called fracton models~\cite{nandkishore2019_fractons,pretko2020_fracton,chamon2005_glass,haah2011_code,yoshida2013_fractal,vijay2015_topo,vijay2016_fractopo,pretko2017_subdim,pretko2018_gaugprinciple,pretko_witten,williamson2019_fractonic}. These are characterized by elementary excitations with restricted mobility (the fractons), whereas non-trivial dynamics can be carried by multi-fracton composites. Recently, fractonic systems conserving both a global $U(1)$ charge as well as its associated dipole moment have successfully been implemented in cold atomic quantum simulation platforms via the application of strong linear potentials~\cite{guardado-sanchez:2020,kohlert:2021,scherg:2021,zahn:2022}. In this context, much effort -- both in theory and experiment -- has been devoted to uncovering the highly exotic nonequilibrium properties of fractonic systems with dipole conservation. These range from dynamical localization~\cite{sala:2020,khemani:2020,rakovszky:2020,moudgalya2021_krylov_me,moudgalya2022_commutant,kohlert:2021,scherg:2021} over novel hydrodynamic universality classes~\cite{gromov:2020,feldmeier:2020,morningstar2020_kinetic,zhang_2020,iaconis2021_multipole,Glorioso:2022ut,grosvenor2021_hydro,osborne2022_fracton,burchards:2022,guardado-sanchez:2020,Iaconis19,feldmeier2021_fractondimer} and glassy dynamics~\cite{chamon2005_glass,prem2017_glassy} to unconventionally slow spreading of quantum information~\cite{feldmeier:2021,hahn2021_information}.

Less attention has been devoted to understand the ground states of fractonic systems. Nonetheless, a gapless Luttinger liquid has been identified as ground state in certain strongly fragmented dipole-conserving spin chains~\cite{rakovszky:2020}. Furthermore, a recent duality mapping between fracton gauge theories and elasticity theory~\cite{pretko2018_elasticity,gromov_elasticity,kumar2019_frac,zhai2019_2dmelting,radzihovsky:2020,zhai:2021} suggests the possible existence of new phases with highly unconventional properties, such as dipole superfluids or fracton condensates~\cite{pretko2018_elasticity,PhysRevLett.121.235301,PhysRevB.100.134113,yuan2020_dSF,chen2021_dSF,radziohvsky2022_lifshitz}. Similar phases have recently also been predicted in a mean-field study of a Bose-Hubbard lattice model subject to dipole conservation~\cite{lake:2022}. However, in one spatial dimension, where generically quantum fluctuations are expected to be strong, an understanding of the phases and phase transitions has been lacking so far. 

In this paper, we address this challenge by studying the Bose-Hubbard model with dipole conservation in one spatial dimension. The one-dimensional character of the system enables us to employ an established toolbox of efficient theoretical techniques.
On the one hand, we resolve the question of a consistently-defined local dipole density, which  subsequently allows us to use bosonization~\cite{giamarchi2003quantum} for constructing effective low-energy field theories of the fracton model. 
On the other hand, we apply tensor network techniques as efficient numerical tools for the computation of ground-state properties of one-dimensional systems~\cite{MurgReview,schollwock:2011}. 

The microscopic model we focus on throughout this paper consists of interacting lattice bosons on a chain subject to the conservation of both charge (i.e., the boson particle number) and dipole moment (i.e., the boson center of mass). In such a \textit{constrained} Bose-Hubbard model the single particle hopping term is absent and is instead replaced by symmetric correlated hopping processes of two bosons. Our microscopic model is described by the Hamiltonian
\begin{equation} \label{eq:1.1}
    \begin{split}
        \hat{H} = -t \sum_{j} (\cre{\hat{b}}{j}\ann{\hat{b}}{j+1}\ann{\hat{b}}{j+1}\cre{\hat{b}}{j+2} + \text{H.c.}) \\
        +\frac{U}{2}\sum_{j} \hat{n}_{j}(\hat{n}_{j}-1) - \mu\sum_{j} \hat{n}_{j}.
    \end{split}
\end{equation}
Here, $t$ denotes the dipole hopping amplitude, $U$ the strength of on-site interactions, $\mu$ the chemical potential, and $\hat{n}_j = \cre{b}{j}\ann{b}{j}$ the local boson number operator. 
Both the total charge $\hat{Q}$ (or particle number $\hat{N}$) and its associated dipole moment $\hat{P}$ are conserved quantities, which we define as
\begin{equation} \label{eq:1.2}
\begin{split}
    \hat{Q} &= \sum_{j=1}^L \hat{q}_j = \sum_j (\hat{n}_j - n) = 0 \\ 
    \hat{P} &= \sum_{j=1}^L (L-j)\, \hat{q}_j = \sum_j (L-j)\, (\hat{n}_j-n) = \mathrm{const.},
\end{split}
\end{equation}
where $\hat{q}_j$ denotes the local deviation from the average boson density $n=\braket{\hat{n}}$. Selecting the reference position of the dipole moment as in \eq{eq:1.2} will turn out convenient in the following. We introduce the notation of a dipole operator $\cre{d}{j} = \cre{b}{j}\ann{b}{j+1}$, such that the kinetic term $\cre{d}{j}\ann{d}{j+1}$ may be viewed as regular nearest-neighbor hopping for a particle-hole dipole-like degree of freedom. We emphasize, however, that the $\hat{d}^{(\dagger)}_j$ do \textit{not} satisfy the commutation relations of creation/annihilation operators. Accordingly, $\hat{d}^\dagger_j\hat{d}_j$ is in general not the local dipole density. However, under certain circumstances it can be, such as in the low-energy subspace considered in Ref.~\cite{sachdev:2002}. Longer range correlated kinetic terms may in principle be included and should not qualitatively affect the low-energy physics. In our numerical computations we restrict ourselves to the simplest case of \eq{eq:1.1}.

\begin{figure}
    \includegraphics[width=\columnwidth]{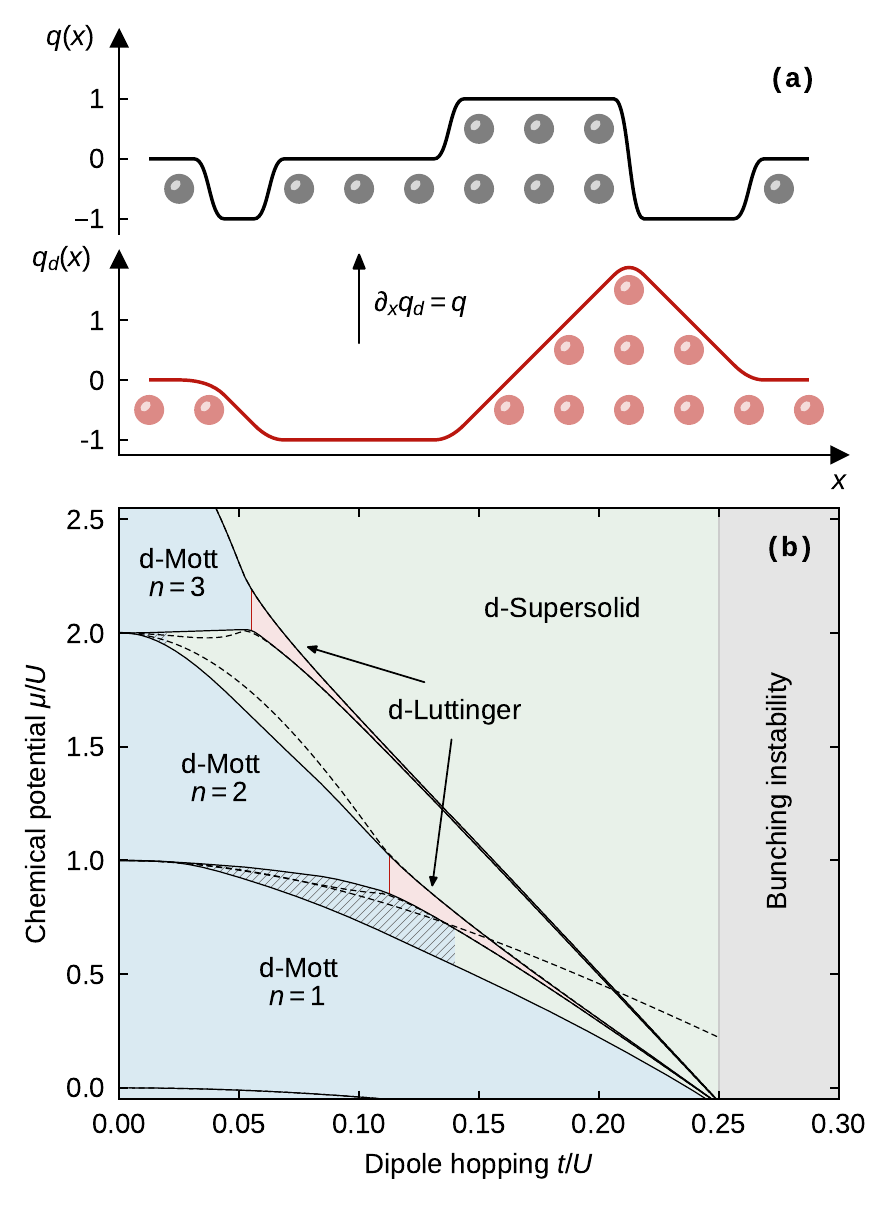}
    \caption{\label{fig:1}
        \textbf{Fractonic phases of matter in one dimension.}
        (a)~At low energies, area law fluctuations of the charge $q(x)$ permit the definition of a local dipole density $q_d(x)$ as $\partial_x q_d(x) = q(x)$.
        This allows us to apply bosonization to construct a low-energy effective field theory for microscopic dipoles, which are composites of fractons.
        (b)~The grand-canonical phase diagram of the dipole-conserving Bose-Hubbard model features three distinct phases: an incompressible dipole Mott insulator (d-Mott), shown in blue, within lobes of integer filling; an incompressible dipole condensate in form of a Luttinger liquid of dipoles (d-Luttinger), located in the red region at the tips of lobes which extends to the bunching instability (grey region); and a compressible supersolid of dipoles (d-Supersolid) at non-integer filling in the green region.
        Solid black lines correspond to estimated phase boundaries from grand-canonical iDMRG computations.
        The dashed black lines indicate the energies for adding or removing a single particle (see text below).
        The regions between the Mott lobes at small dipole hopping $t/U$ (hatched region) additionally host a Mott insulating phase at non-integer filling, which for instance at $n=3/2$ is stable up to $t/U\approx0.14$.
    }
\end{figure}

Our analysis of the zero-temperature phases of \eq{eq:1.1} yields several key results, which we present as follows. In Sec.~\ref{sec:I}, we first establish the presence of area-law cumulative charge fluctuations as a general criterion for the existence of a consistently defined local dipole density; see \figc{fig:1}{a} for an illustration. Using an explicit mapping to microscopic dipole degrees of freedom, we determine the ground-state phases of the model \eq{eq:1.1} at integer boson filling as a function of correlated hopping strength $t/U$ in Sec.~\ref{sec:commensurate}. We predict that the system undergoes a BKT (Berezinskii-Kosterlitz-Thouless) transition between a dipole Mott insulator (d-Mott) and a dipole Luttinger liquid (d-Luttinger). In the dipole Mott insulator both charges and dipoles are gapped, whereas in the dipole Luttinger liquid dipoles are gapless but charge excitations retain a finite energy gap. The dipole Luttinger liquid persists when increasing $t/U$ up until an instability towards boson bunching occurs. We confirm these analytical predictions numerically using large-scale density matrix renormalization group (DMRG) calculations. As a next step, we consider the model away from integer filling in Sec.~\ref{sec:incommensurate}. Our numerical analysis in this regime is consistent with an exotic ground state with vanishing charge gap and thus finite compressibility, described by a quantum Lifshitz model (see e.g.~\cite{seiberg2022_lifshitz}). 
This state spontaneously breaks the continuous dipole symmetry, which, as has recently been shown, is allowed in principle even in one dimension, due to a modified Mermin-Wagner theorem in systems with multipole conservation laws~\cite{PhysRevB.105.155107,kapustin2022_mermin}. In Ref.~\cite{lake:2022}, the quantum Lifshitz model was proposed as low-energy effective theory for the constrained Bose-Hubbard model in a phase termed ``Bose Einstein insulator''. In our one-dimensional scenario, we demonstrate that this state is characterized by a coexistence of density-wave order and dipole superfluidity. We thus refer to this situation as a ``dipole supersolid'' (d-Supersolid). Generic theoretical arguments suggest that the dipole supersolid will eventually become unstable in the thermodynamic limit due to lattice effects. Nonetheless, the full consistency of our results with a dipole supersolid phase within all numerically accessible system sizes demonstrates that the phenomenology of the dipole supersolid is remarkably robust. Our results can be summarized in the phase diagram of \figc{fig:1}{b}. We conclude in Sec.~\ref{sec:outlook} with a discussion of the implications of our results for potential future experimental and theoretical investigations.

\section{Constructing a local dipole density}\label{sec:I}
The ground-state phases studied in this paper require the existence of a bounded local density of microscopic dipoles. This property will be instrumental for us in devising an appropriate low-energy description for the model \eq{eq:1.1}. Such a local dipole density can be seen as an emergent property whose definition is consistent only at low energies and does not extend to high energy states of such dipole-conserving systems.
In the following, we express the conserved global dipole moment in terms of a local density that will remain bounded if charge fluctuations can be shown to be bounded. The most natural way to satisfy this criterion is the presence of a finite charge gap, corresponding to an incompressible state. In such a scenario, the low-energy theory of the system is naturally given in terms of effective dipole degrees of freedom as described in~\cite{PhysRevB.105.155107}. Here, we show how this applies even to a microscopic description of the system.

\subsection{In the continuum}
Let us first consider the scenario of a continuum charge density $q(x)$ in a closed system of length $L$. We require both the total charge and the associated dipole moment to be conserved,
\begin{equation} \label{eq:2.1}
    \begin{split}
        Q &= \int_0^L dx\, q(x) = 0, \\
        P &= \int_0^L dx\, (L-x)\, q(x) = \mathrm{const.}
    \end{split}
\end{equation}
Here, $q(x)=n(x)-n$ denotes again the deviation of the local particle density $n(x)$ from the average density $n$. Our goal is to express the dipole moment as $P=\int_0^L dx\, q_d(x)$ in terms of a local and bounded dipole charge density $q_d(x)$. We emphasize that the naive choice $q_d(x) = x q(x)$ suggested by \eq{eq:2.1} is not suitable since $x\, q(x)$ is manifestly unbounded. Instead, we can use Cauchy's formula for repeated integration to rewrite the dipole moment as
\begin{equation} \label{eq:2.2}
    P = \int_0^L dx\, (L-x)\, q(x) = \int_0^L dx \int_0^x dx^\prime \, q(x^\prime),
\end{equation}
Based on \eq{eq:2.2} we define the local dipole charge density as
\begin{equation} \label{eq:2.3}
    q_d(x) = \int_0^x dx^\prime \, q(x^\prime),
\end{equation}
or alternatively, in differential form,
\begin{equation} \label{eq:2.4}
    \partial_x q_d(x) = q(x).
\end{equation}
The field $q_d(x)$ is thus related to a ``height field'' representation of the dipole constraint~\cite{moudgalya2021_spectral}.
We now see that while $x q(x)$ is unbounded, $q_d(x)$ defined in \eq{eq:2.3} remains bounded if the charge fluctuations within a region of size $x$ remain of order $\mathcal{O}(1)$ as $x \rightarrow \infty$.
As the fluctuations do not scale with the ``volume'' $x$ of the region but originate solely from its boundaries, we will refer to these fluctuations as ``area-law'' in the following.  
Such area-law-type charge fluctuations are guaranteed for the ground state in the presence of a finite charge gap, which induces a finite correlation length for charged degrees of freedom. We therefore obtain a consistently defined local dipole density upon which we can construct an effective model of the low-energy behavior.

\subsection{On the lattice}
The description of the system in terms of a finite density of microscopic dipole charges introduced in \eq{eq:2.4} can also be realized on a lattice. For this purpose, we substitute the continuum derivative with a discrete lattice derivative, $\Delta_x q_{d} := q_{d,x+1/2} - q_{d,x-1/2}$. We have thus defined the local dipole charge as a local bond degree of freedom. 

\begin{figure}
    \includegraphics[width=\columnwidth]{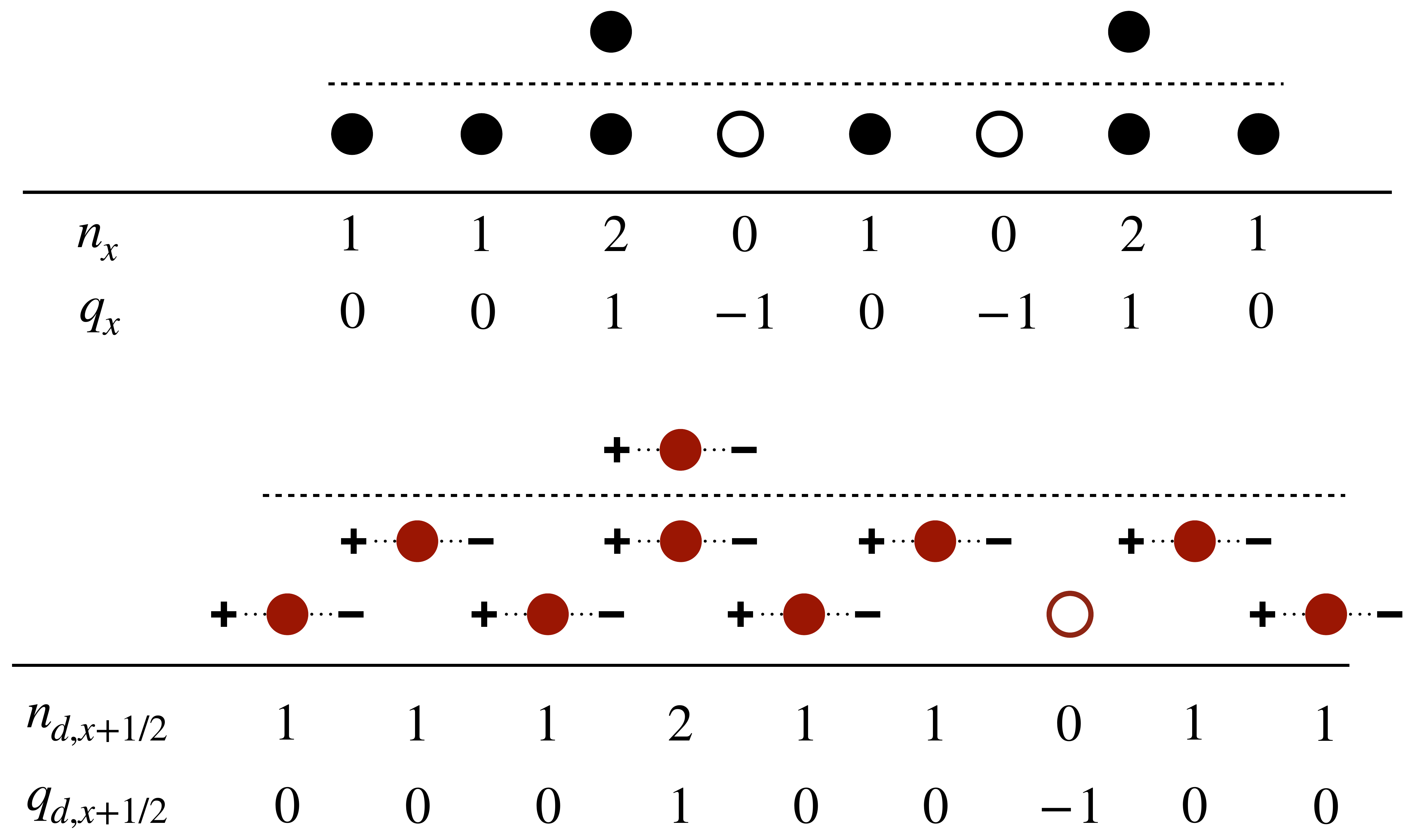}
    \caption{\label{fig:mapping}
        \textbf{Microscopic dipole density.} Mapping between product states in the boson occupation number basis (upper panel) and microscopic dipole occupation numbers on the bonds of the lattice (lower panel). Here, $n_x$ and $n_{d,x+1/2}$ are non-negative, whereas $q_x$ and $q_{d,x+1/2}$ are defined with respect to the average densities (dashed line). For a state at integer boson filling within the same dipole moment sector as the uniform state $\ket{n}$, the resulting dipole model exhibits integer filling as well. See main text for a detailed description of the mapping.
    }
\end{figure}

For simplicity, we focus on integer filling $n \in \mathds{N}$, where any occupation number basis state $\ket{\bs{n}} = \ket{n_1,...,n_L}$ gives rise to a charge density $\ket{\bs{q}} = \ket{n_1-n,...,n_L-n}$ in terms of the local deviation from average filling. The corresponding local dipole charge density state $\ket{\bs{q_d}} = \ket{q_{d,3/2},...,q_{d,L-1/2}}$ can thus be obtained by sweeping through the system from left to right and applying the relation
\begin{equation} \label{eq:2.5}
    q_{d,x+1/2} = q_{d,x-1/2} + q_x,
\end{equation}
where we for now set $q_{d,1/2}=0$. The so-defined local dipole charge can assume both positive and negative values. Much like the conventional charge density, we would like to rewrite the local dipole charge in terms of a non-negative local occupation number $n_{d,x+1/2}$ of microscopic dipoles. This can be achieved simply by adding a suitable integer constant $m \in \mathds{N}$ to the local dipole charge
\begin{equation} \label{eq:2.6}
    n_{d,x+1/2} = q_{d,x+1/2} + m = n_{d,x-1/2} + q_x,
\end{equation}
where now $n_{d,1/2} = m$. Note that the addition of such a constant leaves the differential relation \eq{eq:2.4} invariant. The constant $m$ can be chosen arbitrarily, and we obtain non-negative local dipole occupation numbers $n_{d,x+1/2} \geq 0$ for all $x$ when 
\begin{equation} \label{eq:2.7}
    m \geq m_{\mathrm{min}} = -\min\biggl\{0, \min_{x}\bigl\{q_{d,x+1/2}\bigr\} \biggr\}.
\end{equation}
An illustration of the mapping between $n_x$ and $n_{d,x}$ is provided in \fig{fig:mapping}.
We emphasize that in the presence of a finite charge gap the local dipole charge is always of order $\mathcal{O}(1)$, and thus the required $m_{\mathrm{min}}$ in \eq{eq:2.7} remains bounded as well.

The mapping between boson occupation numbers and bounded dipole occupation numbers can in principle also be performed for states at non-integer boson fillings, provided the charge fluctuations are bounded. In such a case, however, the dipole density \eq{eq:2.3} is defined  with respect to a \textit{nontranslationally invariant} reference state $n_0(x)$, such that $q(x)=n(x)-n_0(x)$. The resulting model for microscopic dipoles is then not translationally invariant. It is an interesting open question how an analysis of such a model can prove useful.
Formally, the mapping could even be performed for arbitrary states $\ket{\bs{n}}$ in the Hilbert space. However, for most states this will lead to an unbounded local dipole density that diverges with system size. The presence of a finite charge gap then ensures that only such occupation number basis states that yield a bounded local dipole density contribute significantly to the ground-state wave function. The contribution of states requiring high local dipole density decays exponentially with $m_{\mathrm{min}}$ and can thus be safely discarded. Furthermore, while the presence of a finite charge gap is a sufficient condition to ensure area-law charge fluctuations, it is not a necessary one.
We will encounter such a situation in Sec.~\ref{sec:incommensurate} in which the charge gap vanishes but cumulative charge fluctuations obey an area law.
We further emphasize that the resulting description in terms of microscopic dipole bond degrees of freedom remains valid for dipole-conserving systems with longer-range terms than in the present microscopic model \eq{eq:1.1}.

\begin{figure*}
    \includegraphics[width=\textwidth]{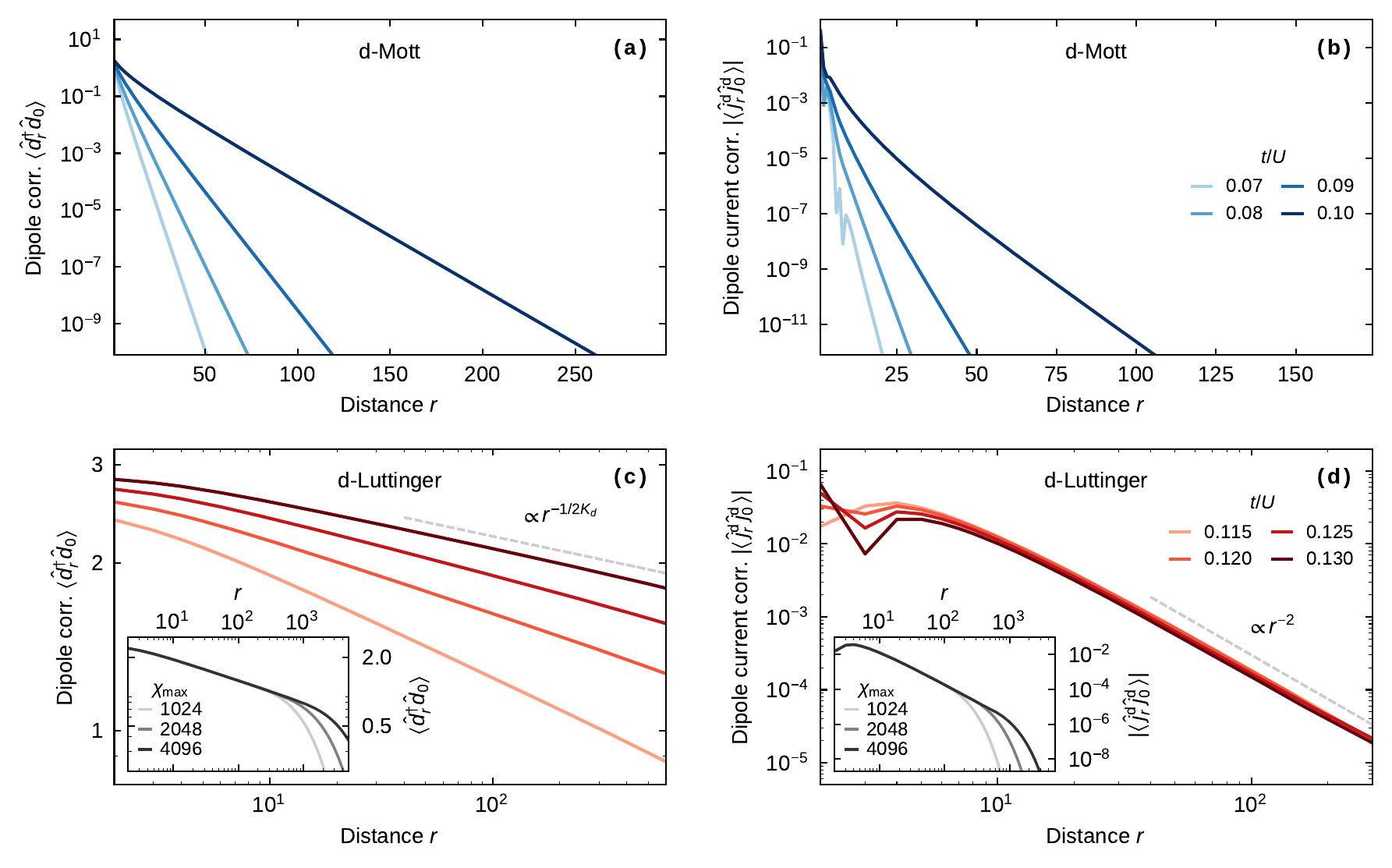}
    \caption{\label{fig:3}
        \textbf{Decay of spatial correlation functions at integer filling.}
        We probe dipole and dipole-current correlations at fixed integer filling $n=2$. [(a), (b)]
        In the Mott insulating phase ($t<t_\mathrm{BKT}$), dipole correlations and dipole-current correlations decay exponentially.
        (c)~Dipole correlations in the Luttinger liquid phase ($t>t_\mathrm{BKT}$) show a power-law decay with the non-universal exponent $1/2K_d$, where $K_d$ is the dipole Luttinger parameter.
        (d)~Dipole-current correlations decay universally with the square of the distance $\propto r^{-2}$.
        The data are obtained with iDMRG, and the insets in (c) and (d) depict the convergence of the correlation functions with bond dimension for $t/U=0.115$ towards the power law decay.
    }
\end{figure*}

\section{Integer filling: Low-energy dipole theory} \label{sec:commensurate}
We start our analysis of the constrained Bose-Hubbard model of \eq{eq:1.1} by considering the system at a fixed integer filling $n\in \mathds{N}$ as a function of the relative strength $t/U$ of the correlated hopping. For $t/U$ being sufficiently small, we expect a Mott insulating state with gapped charge (i.e., single particle) excitations. We then perform the mapping to a system of microscopic dipoles and construct a low-energy effective theory by bosonization of these lattice dipoles.

\subsection{Effective action of dipoles}
In order to determine the proper low-energy model in the dipole language, we extract the resulting average dipole density $n_d$ that results at integer boson filling $n\in \mathds{N}$. In particular, in the following we fix the sector of the total dipole moment $P=0$ that is associated with the homogeneous boson state $\bs{n}=\ket{n,...,n}$. For this state, the local deviation from the average boson filling is $q_x=0$ for all $x$, and therefore the local deviation from the average dipole filling is $q_{d,x+1/2}=0$ for all $x$ as well according to \eq{eq:2.5}. As a result of \eq{eq:2.6}, the average dipole density is thus given by
\begin{equation} \label{eq:2.8}
    n_d = m \in \mathds{N},
\end{equation}
i.e., microscopic dipoles are at integer filling as well. This feature will become relevant upon constructing an appropriate low-energy theory. We emphasize that the states in the sector connected to the homogeneous root state $\ket{\bs{n}}=\ket{n,...,n}$ are obtained by simple hopping processes of the microscopic dipoles, and thus feature the same integer dipole filling.

The presence of a charge gap allows us to rewrite the constrained Bose-Hubbard model at integer boson filling in terms of microscopic bond dipoles at integer filling $n_d \in \mathds{N}$. The Hamiltonian may then be expressed in this basis, leading to a hopping of bond dipoles as well as dipole density interactions. In order to understand the low-energy properties of this system we may then proceed by standard bosonization~\cite{giamarchi2003quantum} of the newly found dipole objects. In particular, we introduce a counting field $\phi_d$ for the bond dipoles, in terms of which the local dipole density reads
\begin{equation} \label{eq:2.9}
    n_d(x) = \Bigl[n_d - \frac{1}{\pi}\nabla \phi_d(x)\Bigr] \sum_p e^{2ip(\pi n_d x -\phi_d(x))}.
\end{equation}
We further introduce a conjugate dipole phase field $\theta_d$, which satisfies the relation
\begin{equation} \label{eq:2.10}
    \bigl[\frac{1}{\pi}\nabla\phi_d(x),\theta_d(x^\prime)\bigr] = -i \delta(x-x^\prime).
\end{equation}
The low-energy effective Hamiltonian for the system is generically given by the kinetic energy $(\nabla \theta_d)^2$ as well as the dipole density interactions $(\nabla \phi_d)^2$. Crucially, since the dipole filling $n_d$ is integer with respect to the original lattice spacing, a cosine term $\cos\bigl(2\phi_d(x)\bigr)$ induced by the underlying lattice needs to be included. Accordingly, the effective Hamiltonian is
\begin{equation} \label{eq:2.11}
\begin{split}
H = \frac{1}{2\pi}\int dx\, \Bigl\{ \frac{u_d}{K_d}(\nabla \phi_d(x))^2 &+ u_dK_d(\nabla \theta_d(x))^2 + \\
&+ g \cos\bigl(2\phi_d(x)\bigr) \Bigr\},
\end{split}
\end{equation}
with the dipole Luttinger parameter $K_d$ as well as the velocity $u_d$. The corresponding Lagrangian for the $\phi_d$ field then reads
\begin{equation} \label{eq:2.12}
    \mathcal{L} = \frac{1}{2\pi K_d} \Bigl\{\frac{1}{u_d} (\partial_\tau \phi_d)^2 + u_d (\partial_x \phi_d)^2\Bigr\} + g \cos\bigl(2\phi_d\bigr).
\end{equation}

\subsection{Dipole Mott insulator to dipole Luttinger liquid transition}
The model \eq{eq:2.12} constitutes the standard low energy theory for interacting lattice bosons at integer filling, and can thus be treated in complete analogy to the usual Bose-Hubbard model.
In particular, the ground state of the model \eq{eq:2.12} undergoes a BKT transition between a gapped Mott insulating phase and a gapless Luttinger liquid at a critical value 
\begin{equation} \label{eq:2.13}
    K_d^* = 2
\end{equation}
of the dipole Luttinger parameter. Above this value the cosine term becomes irrelevant and the system enters a Luttinger liquid of dipoles.
Accordingly, only correlations of the dipole variables $\phi_d$, $\theta_d$ decay algebraically at long distances in the dipole Luttinger liquid. In particular, the vortex operators $e^{i\theta_d(r)}$ that create a dipole at position $r$ decay asymptotically for large distances as
\begin{equation} \label{eq:2.14}
    \braket{e^{i\theta_d(r)}e^{-i\theta_d(0)}} \sim |r|^{-1/2K_d}.
\end{equation}
We will use the characteristic algebraic decay of these correlations in the following to numerically verify the above prediction of a $K_d^*=2$ transition between a dipole Mott insulator (d-Mott) state and a dipole Luttinger liquid (d-Luttinger).  We emphasize that while dipole excitations become gapless, charged particle excitations retain a finite energy gap in the dipole Luttinger liquid. 

We use tensor network techniques to numerically study the ground state phase diagram of our microscopic model \eqref{eq:1.1} at integer boson filling.
Matrix product states (MPS) allow us to obtain an unbiased variational approximation to the many-body ground-state wave function, utilizing the well-established density matrix renormalization group (DMRG) algorithm~\cite{white:1992,schollwock:2011,MurgReview}. Formally, the local Hilbert space of bosons is infinite. In our numerical simulations we impose a cutoff of $n_\text{max} = 8$ particles. 
While MPS are an efficient representation for one-dimensional gapped system, gapless phases such as the expected dipole Luttinger liquid pose a significant numerical challenge.
To best utilize the numerical technique, we implemented both U(1) particle number conservation and dipole conservation~\cite{zaletel:2013} in our DMRG algorithm, enabling us to perform simulations with high bond dimensions. Resolving dipole conservation in our DMRG approach further allows us to numerically determine the energy gap of dipole-like particle-hole excitations. In addition, in order to eliminate the boundary effects of finite systems we will work directly in the thermodynamic limit using infinite DMRG (iDMRG) whenever suitable~\cite{vidal:2007}. A detailed description of our numerical approach is provided in the Appendix.

\textit{\textbf{Dipole and dipole-current correlations.---}}
A direct signature of the transition between a Mott state and a dipole Luttinger liquid is provided by the dipole correlations of \eq{eq:2.14}. These decay exponentially in the Mott phase and algebraically, as in \eq{eq:2.14}, in the Luttinger liquid. 
We probe these correlations numerically in iDMRG by computing
\begin{equation} \label{eq:2.19}
    \braket{\hat{d}^\dagger_r\hat{d}_0} \sim \braket{e^{i\theta_d(r)}e^{-i\theta_d(0)}},
\end{equation}
which is proportional to the correlation of vortex operators $e^{i\theta_d(r)}$ that locally create dipoles.  \figc{fig:3}{a,c} demonstrate that such dipole correlations indeed change from an exponential decay in the Mott insulating phase for $t<t_\text{BKT}$ to power law decay for $t>t_\text{BKT}$. We determine the numerical value of the transition point $t_\text{BKT}$ below. As can be inferred from \figc{fig:3}{c}, the exponent of the power law changes with hopping $t$ and is thus non-universal as expected for a Luttinger liquid.

Besides the dipole correlations, a clear signature of the Luttinger liquid can be obtained by probing the correlations of the dipole current $i\nabla \theta_d(r)$, which in the dipole Luttinger liquid decay at long distances as
\begin{equation} \label{eq:2.20}
    \braket{i\nabla \theta_d(r) i\nabla \theta_d(0)} \sim \frac{1}{2K_d\,r^2}.
\end{equation}
Thus, their power law is independent of the Luttinger parameter $K_d$. Within our microscopic model, the dipole current can be defined and evaluated numerically via the operators
\begin{equation} \label{eq:2.21}
    \hat{j}^{d}_{j} = -i \bigl(\cre{d}{j}\ann{d}{j+1} - \mathrm{H.c.}\bigr).
\end{equation}
Our numerical results in \figc{fig:3}{b,d} show that correlations $\braket{\hat{j}^d_r\hat{j}^d_0}$ of this dipole current decay exponentially in the Mott state for $t<t_\text{BKT}$ and indeed fall off as the inverse square of the distance $r$ for $t>t_\text{BKT}$. The slight vertical shift of the corresponding curves in \figc{fig:3}{d} is nonuniversal and depends on the Luttinger parameter $K_d$.

\begin{figure}
    \includegraphics[width=\columnwidth]{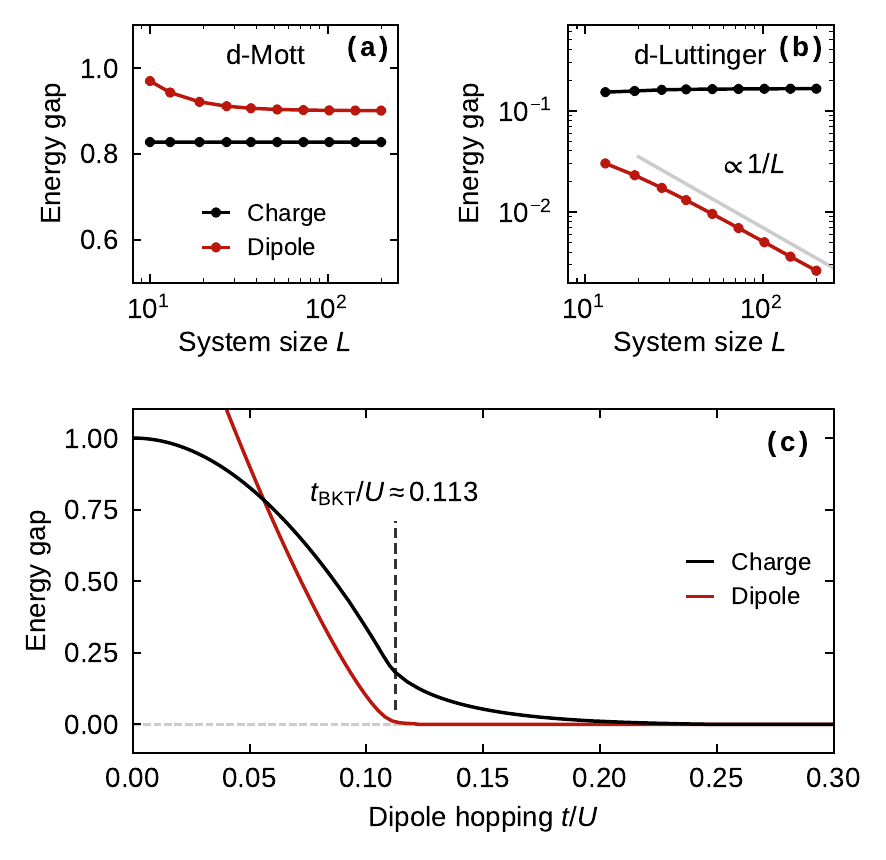}
    \caption{\label{fig:gap}
        \textbf{Energy gaps.}
        Finite size flow of the excitation gaps at integer filling $n=2$ (a) in the Mott insulator ($t/U=0.050$) and (b) the Luttinger liquid ($t/U=0.115$).
        Both in the Mott and Luttinger liquid phase the charge gap converges to a finite value $\Delta_c \to \text{const.}$ as $L\rightarrow\infty$.
        By contrast the dipole gap remains finite only in the Mott insulator, but vanishes in the Luttinger liquid as $\Delta_d\propto 1/L$.
        (c)~Charge and dipole excitation gap across the BKT transition. The dipole gap closes at the critical  hopping $t_\mathrm{BKT}/U \approx 0.113$.
    }
\end{figure}

\begin{figure}
    \includegraphics[width=\columnwidth]{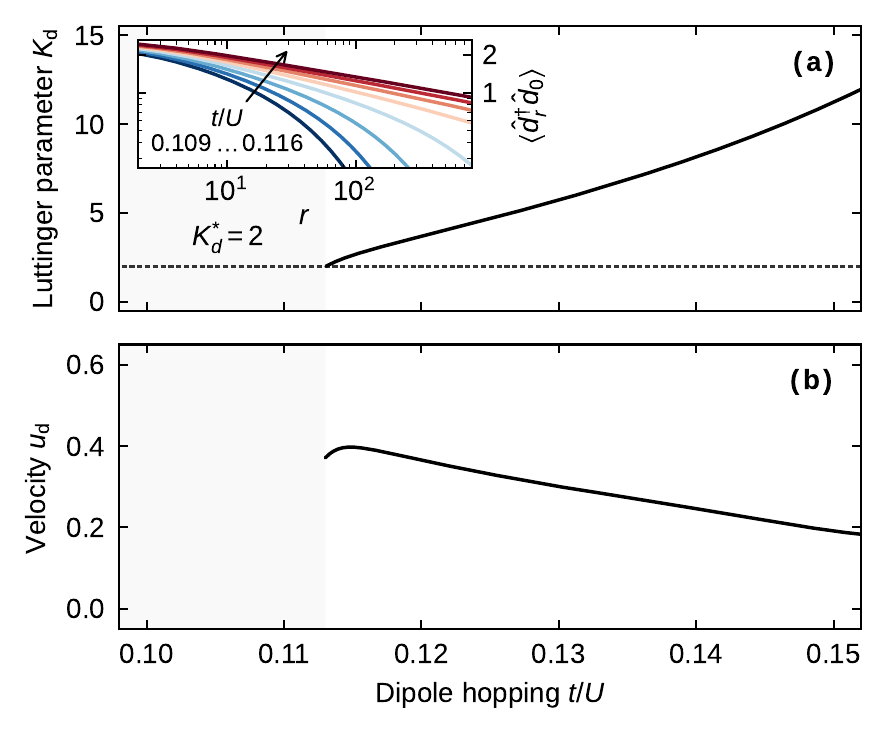}
    \caption{\label{fig:4}
        \textbf{Characterization of the dipole Luttinger liquid at commensurate filling $n=2$.}
        (a)~Luttinger parameter extracted from the asymptotics of the dipole correlations $\langle \hat{d}^{\dagger}_{r}\hat{d}^{\phantom\dagger}_{0} \rangle \propto r^{-1/2K_d}$. 
        The inset demonstrates the transition from exponential to power-law decay of the dipole correlations around the critical point $t_\mathrm{BKT}$, which occurs precisely at the predicted value of the Luttinger parameter $K_{d}^{*}=2$.
        (b)~Velocity $u_d$ obtained from the dipole compressibility $\kappa_d = K_d/ u_d \pi$.
    }
\end{figure}

\textit{\textbf{Energy gaps.---}}
In the Mott insulating phase, both particle excitations and dipole excitations feature a finite energy gap. The transition to the dipole Luttinger liquid should be accompanied by a closing of the dipole gap while the gap for charged particle excitations remains finite. Our numerical approach allows us to explicitly verify these expectations.

Let us consider the system at some integer boson filling $N=nL$ and a dipole moment $P=0$ that corresponds to the one of the homogeneous state $\ket{n,...,n}$ see \eq{eq:2.1}. The filling $n$ can be thermodynamically stable when the chemical potential $\mu$ in \eq{eq:1.1} is located between the two potentials
\begin{equation} \label{eq:2.15}
    \begin{split}
        \mu_{\text{c}}^{+}(L) &= E_0(L,N+1,P) - E_0(L,N,P), \\
        \mu_{\text{c}}^{-}(L) &= E_0(L,N,P) - E_0(L,N-1,P), \\
    \end{split}
\end{equation}
where $E_0(L,N,P)$ denotes the ground state energy of the system of size $L$ at fixed particle number $N$, dipole moment $P$ and vanishing chemical potential.
Accordingly, as for the conventional Bose-Hubbard model~\cite{Ejima_2011}, the gap to charged single particle excitations in such a system is defined as
\begin{equation} \label{eq:2.16}
\Delta_\text{c}(L) = \mu_{\text{c}}^{+}(L) - \mu_{\text{c}}^{-}(L).
\end{equation}
Analogously, the dipole gap can now be obtained via the two potentials
\begin{equation} \label{eq:2.17}
    \begin{split}
        \mu_{\text{d}}^{+}(L) &= E_0(L,N,P+1) - E_0(L,N,P), \\
        \mu_{\text{d}}^{-}(L) &= E_0(L,N,P) - E_0(L,N,P-1), \\
    \end{split}
\end{equation}
which yields
\begin{equation} \label{eq:2.18}
    \Delta_\text{d}(L) = \mu_{\text{d}}^{+}(L) - \mu_{\text{d}}^{-}(L).
\end{equation}
We notice that $\mu_d^+(L) = - \mu_d^-(L)$ holds, since by spatial reflection symmetry the ground state energy cannot depend on whether a particle-hole excitation is created by displacing a single particle to the right or left.
In the thermodynamic limit, the gaps $\Delta_{c/d} = \lim_{L,N\rightarrow\infty}\Delta_{c/d}(L)$ are obtained by keeping $n=N/L$ and $P=0$ fixed. In our numerical simulations based on iDMRG, we approach this limit by adding/removing a single particle to the unit cell, whose size $L$ is increased until convergence of the gaps is reached. This has the advantage that the system is formally infinite and does not suffer from effects of boundary conditions. In \figc{fig:gap}{a,b}, we show the finite size flow of the charge and dipole gaps both in the Mott insulator and the Luttinger liquid. Both gaps remain finite in the dipole Mott insulator. In the dipole Luttinger liquid the charge gap remains finite, whereas the dipole gap closes as $1/L$.

\figc{fig:gap}{c} shows the numerically determined charge and dipole gaps as functions of correlated hopping $t/U$ that we extrapolate to the thermodynamic limit. We observe a rapid closing of the dipole gap at $t_\text{BKT}/U \approx 0.113$, while at the same time the particle gap remains finite. Our results are thus consistent with a transition from a dipole Mott insulator to a dipole Luttinger liquid at a critical strength of the correlated hopping.

\textit{\textbf{Luttinger parameter and dipole velocity.---}}
In the dipole Luttinger liquid phase, the system is characterized entirely by the value $K_d$ of the Luttinger parameter as well as the dipole velocity $u_d$. For example, we can verify the BKT transition between the Mott state and the dipole Luttinger liquid, which is driven by the cosine term in \eq{eq:2.12}. The BKT theory of this transition predicts a critical dipole Luttinger parameter $K_d^*=2$, which we can verify by extracting $K_d(t)$ as a function of the correlated hopping $t/U$ from the numerically determined dipole correlations \eq{eq:2.19}. \figc{fig:4}{a} shows that the Luttinger parameter continuously increases as a function of $t/U$. The lowest value of the Luttinger parameter $K_d$ is indeed $K_d^* = 2$, which marks the onset of powerlaw dipole correlations. The value of the critical hopping is consistent with the value of $t_\text{BKT}$ obtained from the closing of the dipole gap in \figc{fig:gap}{c}.

Due to the finite charge gap, the dipole Luttinger liquid is incompressible (see also the discussion below). Nonetheless, it features gapless low-energy dipole excitations $\omega = u_d |k|$, with which we associate a finite \textit{dipole compressibility} $\kappa_d$. 
The dipole velocity $u_d$ that we want to extract in order to fully characterize the Luttinger liquid is directly related to this dipole compressibility via $\kappa_{d}=K_d/u_d\pi$.
We use this relation to obtain the dipole velocity by numerically extracting the dipole compressibility from the finite size flow of the dipole gap $\Delta_d(L) = \kappa_{d}^{-1}/L$.
The resulting dipole velocity is shown in \figc{fig:4}{b}. In addition, we have numerically confirmed the existence of linear low energy modes consistent with the estimated velocities of \figc{fig:4}{b} by computing the full dipole spectral function. We will address such dynamical properties in detail in future work.

\begin{figure}
    \includegraphics[width=\columnwidth]{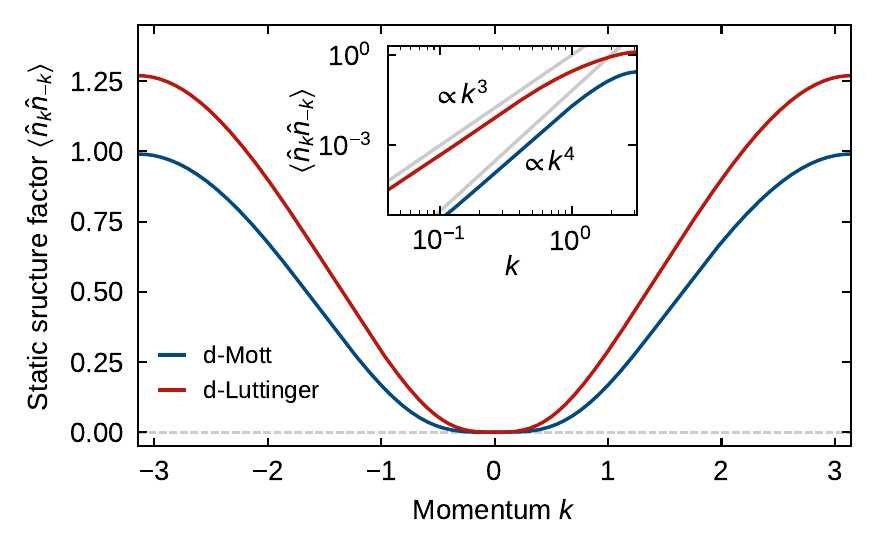}
    \caption{\label{fig:5}
        \textbf{Static charge structure factor.}
        Static structure factor in the  Mott insulator and the Luttinger liquid.
        The inset illustrates the power law decay of $\langle\hat{n}_{k}\hat{n}_{-k}\rangle$ for $k\rightarrow 0$, which is $\propto k^4$ in the Mott phase and $\propto |k|^3$ in the Luttinger liquid, respectively.
    }
\end{figure}

\textit{\textbf{Charge compressibility.---}}
The presence of a finite charge gap guarantees the incompressibility of the dipole Luttinger liquid. Alternatively, the charge compressibility $\kappa$ can be determined via the zero-frequency  density correlations
\begin{equation} \label{eq:2.22}
    \kappa = \lim_{k\rightarrow 0} C_{nn}(\omega=0,k),
\end{equation}
with the structure factor
\begin{equation} \label{eq:2.23}
    C_{nn}(\omega,k) = \braket{n(\omega,k)n(-\omega,-k)}.
\end{equation}
For the dipole Luttinger liquid, the dipole density is given by $n_d \sim \partial_x \phi_d$ according to \eq{eq:2.9} and the corresponding charge density is $n \sim \partial_x^2 \phi_d$ upon using \eq{eq:2.4}. Therefore, the compressibility is 
\begin{equation} \label{eq:2.24}
    \kappa(k) = \frac{1}{\pi^2} k^4 \braket{\phi_d(\omega=0,k)\phi_d(\omega=0,-k)} = \frac{K_d}{u_d\pi} k^2,
\end{equation}
which vanishes as $k^2$ for small momenta. In our DMRG simulations, the frequency-resolved density correlations are challenging to obtain. However, we can efficiently compute the equal-time density correlations $C_{nn}(\tau=0,k)$. For the Luttinger liquid model of \eq{eq:2.12}, the relevant time and frequency correlations are related by
\begin{equation} \label{eq:2.25}
    C_{nn}(\tau=0,k) = \frac{u_d}{2}|k| C_{nn}(\omega=0,k) = \frac{K_d}{2\pi} |k|^3.
\end{equation}
We show the equal-time correlations $C_{nn}(\tau=0,k)$ in \fig{fig:5}, which we numerically obtain from the real space density-density correlations
\begin{equation}
    \braket{\hat{n}_{k} \hat{n}_{-k}} = \frac{1}{L^2} \sum_{j,j'} e^{-ik(j-j')} \braket{\hat{n}_{j}\hat{n}_{j'}}.
\end{equation}
Indeed, we find a $\sim |k|^3$ behavior at small $k$ for the Luttinger liquid in \fig{fig:5}, which in turn is consistent with a compressibility vanishing as $\kappa(k) \sim k^2$. By contrast, in the Mott insulating state with finite dipole excitation gap, the density correlations instead vanish as $C_{nn}(\tau=0,k) \sim k^4$, see \fig{fig:5}.

\subsection{Stability of dipole Luttinger liquid at large correlated hopping}
In the previous section we analyzed the transition out of a gapped Mott state into a gapless dipole Luttinger liquid at integer filling upon increasing the strength $t/U$ of the correlated hopping. It is natural to ask whether a second transition into a state with gapless charge excitations appears as the hopping $t$ is increased even further. A natural candidate for such a phase is the (1+1)D quantum Lifshitz model [see \eq{eq:2.26} below], that has been proposed as a potential theory of gapless phases with dipole-moment conservation.

As we discuss in the following, in the present situation the charge gap remains finite upon increasing $t$. A transition to a phase described by a Lifshitz model does not occur since such a phase is destroyed by lattice effects. This instability of the Lifshitz model can be used to estimate the value of the charge gap at large values of the dipole Luttinger parameter. The dipole Luttinger liquid is therefore stable against a transition into a gapless Lifshitz model.

Nonetheless, for the Hamiltonian of \eq{eq:1.1} the Luttinger liquid will eventually become unstable for $t$ greater than some $t^*$ towards a state in which all bosons bunch together in space. The corresponding ground state features a superextensive energy $E_0 \sim -N^2$ and does not correspond to a stable phase of matter unless a (unphysical) cutoff on the local boson occupation is introduced.

\textbf{\textit{Bunching instability.---}}
The bunching instability can be understood by the fact that both the correlated hopping term and the on-site interaction term scale quadratically with the local occupation number $n$.
For a thermodynamically stable phase of matter, the asymptotic scaling of the ground-state energy for $n\gg 1$ demands $-2tn^2+Un^2/2>0$, hence in a grand-canonical setting the transition occurs precisely at $t^*/U = 0.25$, as for $t>t^*$ the ground state is unstable toward a diverging particle number.
In case of a fixed particle number, however, the situation is somewhat richer. 
At low filling, the reduced local density fluctuations increase the critical value $t^*$.
For $n=1$, we numerically obtain $t^*/U \approx 0.32$, and for $n=2$ we obtain $t^*/U \approx 0.26$. 

\textbf{\textit{Stability of the dipole Luttinger liquid and instability of the Lifshitz model.---}}
For hopping strengths below the bunching instability and at integer filling, the system  remains in the dipole Luttinger liquid and does not enter a phase of gapless charge excitations. Here, we argue why this is the case before determining the asymptotic behavior of the charge gap $\Delta_c$ for large values of the dipole Luttinger parameter. 

Introducing conjugate bosonized variables $\phi(x)$ and $\theta(x)$ for the charge degrees of freedom [analogous to \eqs{eq:2.9}{eq:2.10}], the dipole-conserving yet \textit{charge-gapless} quantum Lifshitz model reads
\begin{equation} \label{eq:2.26}
    \mathcal{L} = \frac{K}{2\pi}\Bigl\{ \frac{1}{v}\bigl(\partial_\tau \theta\bigr)^2 + v\bigl(\partial_x^2 \theta\bigr)^2 \Bigr\}.
\end{equation}
The associated Hamiltonian is given by 
\begin{equation} \label{eq:rev1}
\begin{split}
 H &= \frac{v}{2\pi}\int dx\, \Bigl[\frac{1}{K}(\partial_x\phi)^2 + K(\partial_x^2\theta)^2\Bigr] = \\
 &= \frac{v}{2\pi}\int dk\, \Bigl[\frac{1}{K}k^2|\phi(k)|^2 + Kk^4|\theta(k)|^2\Bigr].
 \end{split}
\end{equation}
In this model the usual kinetic term $(\partial_x \theta)^2$ is quenched and instead a dipole-conserving kinetic term $(\partial_x^2\theta)^2$ invariant under linear shifts $\theta(x) \rightarrow \theta(x) + a + bx$ is the most relevant allowed contribution. The constrained kinetic term induces a relative scaling $z=2$ of space and time coordinates. 

Within an effective field theory approach~\cite{lake:2022}, the quantum Lifshitz model can be obtained upon considering the charge and dipole degrees of freedom as \textit{independent}, coupling them in the total Lagrangian
\begin{equation} \label{eq:2.27}
\begin{split}
    \mathcal{L} = &\frac{K_d}{2\pi} \Bigl[\frac{1}{u_d}(\partial_\tau\theta_d)^2 + u_d(\partial_x \theta_d)^2 \Bigr] + r (\theta_d + \partial_x \theta)^2 + \\
    &+ \frac{K_c}{2\pi u_c} (\partial_\tau \theta)^2,
\end{split}
\end{equation}
and subsequently integrating out the variables $\theta_d$. This theory was first analyzed in the context of a fracton gauge dual formulation of classical smectics in two dimensions~\cite{zhai:2021}.
We emphasize the difference to the microscopic derivation of Sec.~\ref{sec:I}. There, charge and dipole degrees of freedom were not independent but related by a change of variables. In Sec.~\ref{sec:I}, the low energy theory of the dipole Luttinger liquid could be postulated upon \textit{assuming} a finite gap for the charge degree of freedom. As we will see in the following, the benefit of the effective field theory approach of \eq{eq:2.27} is to determine whether/when this assumption can be valid.

In \eq{eq:2.27}, $r>0$ and $K_c/u_c$ quantify the density interaction between charge degrees of freedom. Notice that the term $(\theta_d + \partial_x\theta)^2$ is also invariant under $\theta_d \rightarrow \theta_d - b$, $\theta \rightarrow \theta +a +bx$. Physically, one expects this term to introduce a constraint that induces a finite stiffness for the dipole phase field and pins it to the charge field, $\theta_d \rightarrow -\partial_x \theta$. Formally, after integrating out $\theta_d$, we obtain a Lifshitz model of the form \eq{eq:2.26} with
\begin{equation} \label{eq:2.28}
    K = \sqrt{\frac{u_d K_c}{u_c K_d}}\, K_d, \qquad v = \sqrt{\frac{u_c u_d K_d}{K_c}}.
\end{equation}
We note that $K_c/u_c$ and $K_d/u_d$ quantify the density interactions between charges and dipoles, respectively, both of which derive from the underlying density interaction of the microscopic dipole Bose-Hubbard model. 
We thus naturally expect the ratio $u_dK_c/u_cK_d \approx \mathcal{O}(1)$ in \eq{eq:2.28} to be of order unity, and thus $K \sim K_d$.

We further note that expressed in terms of the $\phi$-field (and in frequency and momentum space),  the Lifshitz model takes the form
\begin{equation} \label{eq:rev2}
\mathcal{L} = \frac{1}{2\pi K}\Bigl[ \frac{1}{v}\frac{\omega^2}{k^2} + vk^2 \Bigr]|\phi(k)|^2.
\end{equation}
This follows from \eq{eq:2.26} and the invariance of the Hamiltonian \eq{eq:rev1} under $K\rightarrow 1/K$, $\theta(k)\rightarrow \phi(k)/k$, $\phi(k)\rightarrow k\theta(k)$. Now if the charge degrees of freedom $\phi$ were to acquire a finite gap, adding a mass term $r(\phi-\partial_x \phi_d)^2$ and taking into account dipole density interactions $\frac{K_d}{2\pi u_d}(\partial_x \phi_d)^2$ in \eq{eq:rev2} returns us to the dipole Luttinger liquid upon integrating out $\phi$. 
Thus, the two effective constraints $\phi-\partial_x\phi_d$ and $\theta_d + \partial_x \theta$ on density and phase variables, driving the system either into the dipole Luttinger liquid or the Lifshitz model, respectively, are in fact conjugate to each other:
\begin{equation}
[\phi(x)-\partial_x\phi_d(x),\theta_d(x^\prime)+\partial_x\theta(x^\prime)] = 2i\pi \delta(x-x^\prime).
\end{equation}

We now show that a finite charge gap is always present in the Lifshitz model due to lattice effects. At integer filling a cosine term
\begin{equation} \label{eq:rev3}
g \cos(2\phi(x))
\end{equation}
for the charge field $\phi$ should be included in our description.
The operators $e^{i\phi(r)}$ have long-range correlations in the model of \eq{eq:rev2} independent of $v$ and $K$. Therefore, the cosine term is always relevant and creates a gap for charged excitations, thus driving the system back into the dipole Luttinger liquid.

\textit{\textbf{Charge gap at large $K_d$.---}}
In the following, we estimate the size of the charge gap at large values of $K_d$ by means of a scaling analysis for local fluctuations of the $\phi$ field. Extracting the charge gap allows us to verify not only that the dipole Luttinger liquid remains stable as the hopping is increased, but importantly also that the mechanism behind the generation of a gap is indeed the presence of a relevant cosine term in the Lifshitz model in \eq{eq:rev2}.

In the presence of a non-zero coupling $g\neq 0$, the cosine is the most relevant term appearing in the action $S = \int d\tau dx \mathcal{L}$ that results from \eq{eq:rev2} and \eq{eq:rev3}. It is thus safe to expand the cosine to quadratic order and consider the model
\begin{equation} \label{eq:2.34}
    \mathcal{L} = \frac{1}{2\pi K}\Bigl[ \frac{1}{v}\frac{\omega^2}{k^2} + vk^2 + 4\pi K g \Bigr]|\phi(k)|^2.
\end{equation}
We emphasize that the two terms $\cos(2\phi)$ and $\phi^2$ indeed have the same scaling dimension in the Lifshitz model.
We now evaluate local correlations of the $\phi$-field within this model yielding
\begin{equation} \label{eq:2.35}
    \begin{split}
        \Braket{\bigl[\phi(\tau=0,x=0)\bigr]^2} = \int_0^{\frac{\pi}{a}} dk \, \frac{K}{\sqrt{1+\frac{4\pi K g}{v k^2}}}, 
    \end{split}
\end{equation}
where we have included a high-momentum cutoff that is set by the microscopic lattice spacing $a$. We see that for $g=0$, the term inside the integral in \eq{eq:2.35} is proportional to $K$. Fluctuations of $\phi$ thus become large as $K$ increases. For nonzero $g\neq 0$ on the other hand, the term inside the integral will eventually become suppressed for sufficiently small momenta $k$, thus reducing fluctuations of $\phi$ on the corresponding length scale. The relevant length scale at which the presence of the cosine becomes noticeable is determined by the momentum at which the term inside the integral in \eq{eq:2.35} is reduced from order $K$ down to order $\mathcal{O}(1)$. Setting $k=1/\lambda$, this leads to the condition
\begin{equation} \label{eq:2.36}
1 \stackrel{!}{=} \frac{K}{\sqrt{1+\lambda^2 \frac{4\pi K g}{v}}} \simeq \frac{K}{\lambda} \sqrt{\frac{v}{4\pi K g}},
\end{equation}
where we have used $\lambda^2 Kg/v \gg 1$ on the relevant length scale $\lambda$ for large values of $K$. The length scale $\lambda$ is thus
\begin{equation} \label{eq:2.37}
\lambda \sim \sqrt{\frac{v K}{g}}.
\end{equation}
Due to the dynamical exponent $z=2$ between space and time in the Lifshitz model, this length scale is associated with a corresponding energy scale
\begin{equation} \label{eq:2.39}
    \Delta_c \sim \lambda^{-2} \sim \frac{g}{vK} = \frac{g}{u_d K_d}.
\end{equation}
In the last step we have inserted the values of \eq{eq:2.28} for $K$ and $v$ that we have derived from the underlying dipole Luttinger liquid. 

\begin{figure}
    \includegraphics[width=\columnwidth]{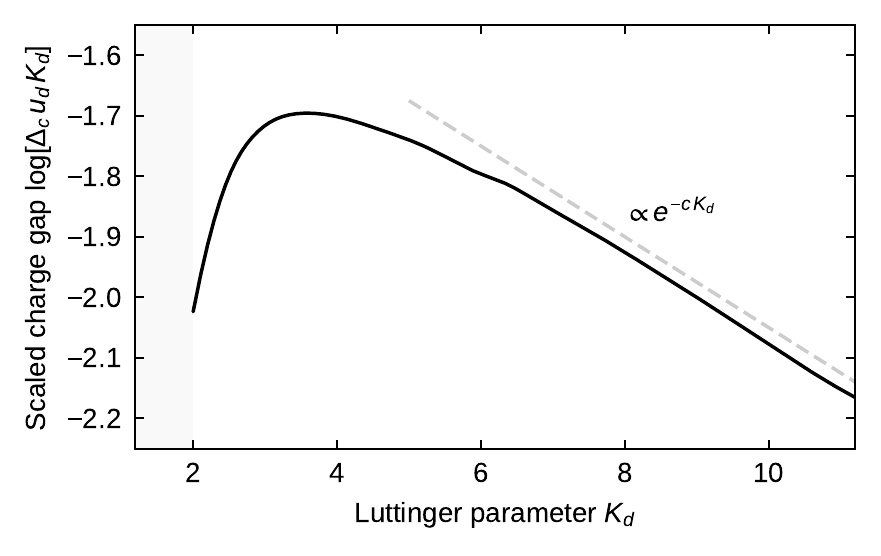}
    \caption{\label{fig:charge_gap}
        \textbf{Charge gap in the dipole Luttinger liquid.}
        Numerically obtained charge gap $\Delta_c$ in the dipole Luttinger liquid, scaled with $u_d\,K_d$, as a function of the Luttinger parameter $K_d$. We compare the scaling of the charge gap to the theoretical prediction $\propto e^{-cK_d}$ with some non-universal constant $c$.
    }
\end{figure}

We have already extracted the dipole Luttinger parameter $K_d(t)$, the dipole velocity $u_d(t)$, and the charge gap $\Delta_c(t)$ as functions of the correlated hopping $t/U$ in our numerics. We can now determine the value $g(t)$ of the cosine term in order to verify the prediction \eq{eq:2.39}. Even though we cannot infer $g(t)$ directly from our numerics, we know that the correlations of the operators $e^{i\phi(r)}$ scale in the Lifshitz model as
\begin{equation} \label{eq:2.40}
    \braket{e^{i\phi(r)}e^{-i\phi(0)}}_{g=0} \xrightarrow{r\rightarrow \infty} e^{-c^\prime K} = e^{-c K_d},
\end{equation}
with non-universal $\mathcal{O}(1)$ constants $c^\prime$, $c$. It is the constant value of this correlation function that turns the cosine term into a relevant operator of the same scaling dimension as a conventional mass term $\phi^2$. As the value \eq{eq:2.40} of this constant becomes small at large $K_d$, the prefactor of the mass term in \eq{eq:2.34} should be small as well, and we thus infer that $g(K_d)$ decays exponentially with $K_d$,
\begin{equation} \label{eq:2.41}
    g(K_d) = c_0\,\exp(-c K_d).
\end{equation}
With \eq{eq:2.41} at hand, we can verify our prediction \eq{eq:2.39} for the charge gap by inverting the relation $K_d(t/U) \rightarrow t/U(K_d)$ and verifying that
\begin{equation} \label{eq:2.42}
    \Delta_c(K_d) u_d(K_d) K_d \sim \exp(-c K_d)
\end{equation}
for large values of $K_d$. In \fig{fig:charge_gap} we display the quantity on the left-hand side of \eq{eq:2.42} calculated from our numerically obtained values for $K_d$, $u_d$, and $\Delta_c$. We indeed find a decay of $\Delta_c u_d K_d$ consistent with an exponential at increasing values of $K_d$, thus confirming \eq{eq:2.39}. We note that the range of available values for $K_d$ in \fig{fig:charge_gap} is limited mostly by the numerical evaluation of the dipole velocity $u_d$~\cite{footnote}, and a larger parameter range would be desirable in order to verify \eq{eq:2.39} more accurately. Interestingly, although the exponential decay of $g(K_d)$ dominates at very large $K_d$, at the available intermediate values of $K_d$ it is essential to take into account the prefactor $1/u_d K_d$ of the gap in \eq{eq:2.39} in order to be able to see the exponential form.

We have thus directly verified that the charge gap -- produced by the instability of the Lifshitz model -- remains finite as the hopping strength is increased towards the bunching transition. The dipole Luttinger liquid thus persists as a stable phase at integer filling.

\section{Non-integer Filling} \label{sec:incommensurate}
In the previous section, we have seen that at integer boson filling, the dipole Luttinger liquid remains stable, and the corresponding charge gap of single-particle excitations stays finite, up to a point at  $t^*/U$ where a bunching instability arises. Naturally, we can ask whether there exists a different parameter regime of the lattice system in which a charge-gapless and thus compressible state described by a Lifshitz model may be realized? In this section, we will explore this question in the regime of non-integer boson fillings.

In particular, let us consider the bosonic lattice model at some rational filling $n=p/q \notin \mathds{N}$, with $p$,$q$ coprime integers. In the putative Lifshitz model of \eq{eq:rev2} and \eqref{eq:rev3} at sufficiently large hopping $t$ (but below bunching) the $\cos(2\phi(x))$ term---which we have previously determined to destabilize the phase at integer filling---is no longer present. Nonetheless, higher order (i.e. multiple) vortex terms in the expansion \eq{eq:2.9} of the density operator may generically still contribute. In particular, for the given filling fraction $p/q$ one may generally expect a contribution 
\begin{equation} \label{eq:3.1}
    g_q \cos(2q\phi(x))
\end{equation}
to the Lagrangian.
Such terms are always relevant and open a charge gap, analogously to the analysis of the previous section. It follows that it should generically be expected that the Lifshitz model is unstable also at any rational filling. Nonetheless, it is possible in principle that the prefactors $g_q$ in \eq{eq:3.1} become either (1) extremely small, such that the phenomenology of the Lifshitz model survives even in very large systems, or (2) exactly zero for some specific microscopic lattice models, such that the Lifshitz phase survives even in the thermodynamic limit. We have no immediate reason to think that the prefactors of such higher-order cosine terms should vanish identically for our model. We note, however, that since all cosine-terms in \eq{eq:3.1} are equally relevant, possible cancellations between different harmonics may occur, potentially generating a situation with effectively very small prefactors.

In the following, we analyze the ground state of the system at non-integer filling numerically using iDMRG. Remarkably, we find that the variational ground state obtained numerically is consistent with a compressible phase described by the Lifshitz model in the absence of any cosine-terms for the accessible system sizes and bond dimensions. We first present evidence for the compressible nature of this variational state before characterizing the physical properties of this phase.
Whether the Lifshitz model will eventually become unstable in regimes beyond our current numerical capacities is an intriguing open question. However, we emphasize that already the observed stability of this phase on our currently accessible scales is quite remarkable and surprising.

\begin{figure}
    \includegraphics[width=\columnwidth]{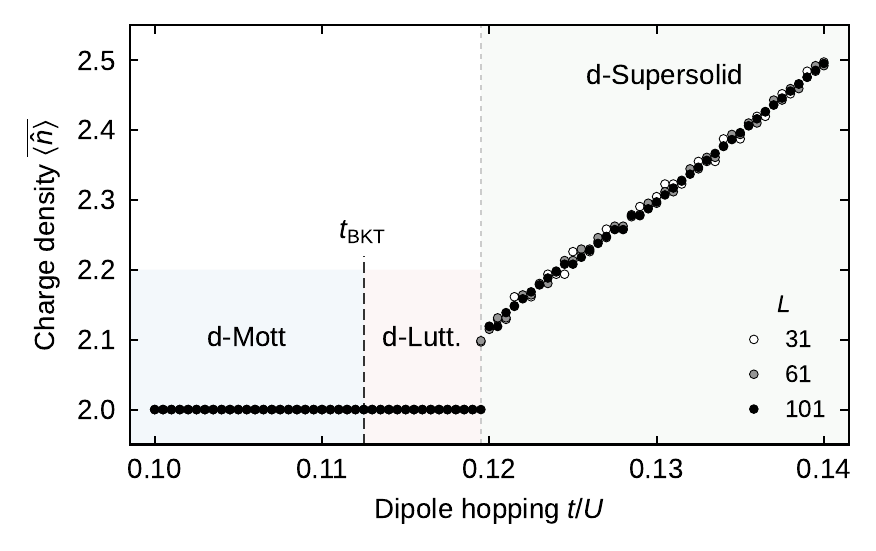}
    \caption{\label{fig:7}\label{fig:non-int_filling}
        \textbf{Transition to incommensurable densities.}
        Average charge density $\overline{\langle \hat{n} \rangle}$ for a cut through the grand-canonical phase diagram along the line of fixed chemical potential $\mu/U=0.95$.
        Different unit-cell sizes $L$ used in the iDMRG simulations are compared. We observe an apparent first-order transition into a compressible state with continuously varying charge density.
    }
\end{figure}

\textit{\textbf{Fixed chemical potential.---}}
We explore non-integer fillings by relaxing both charge and dipole quantum numbers in our numerics and by performing a grand-canonical ground state search as a function of hopping $t/U$ along a line of fixed chemical potential $\mu/U=0.95$. From the previously computed charge and dipole gaps displayed in the phase diagram of \fig{fig:1}, we expect such a line cut to go through the two integer-density phases of the Mott insulator and the dipole Luttinger liquid before reaching a regime of non-integer ground state density. We show the average density expectation value $\braket{\hat{n}}$ along this cut in \fig{fig:7}. Crucially, upon reaching a critical hopping strength, the density appears to exhibit a first-order jump before increasing again continuously. While narrow density-plateaus in the regime of non-integer filling are still visible for smaller unit-cell sizes, these plateaus appear to smoothen out as the unit-cell size is increased, an indication of a compressible state.

\begin{figure}
    \includegraphics[width=\columnwidth]{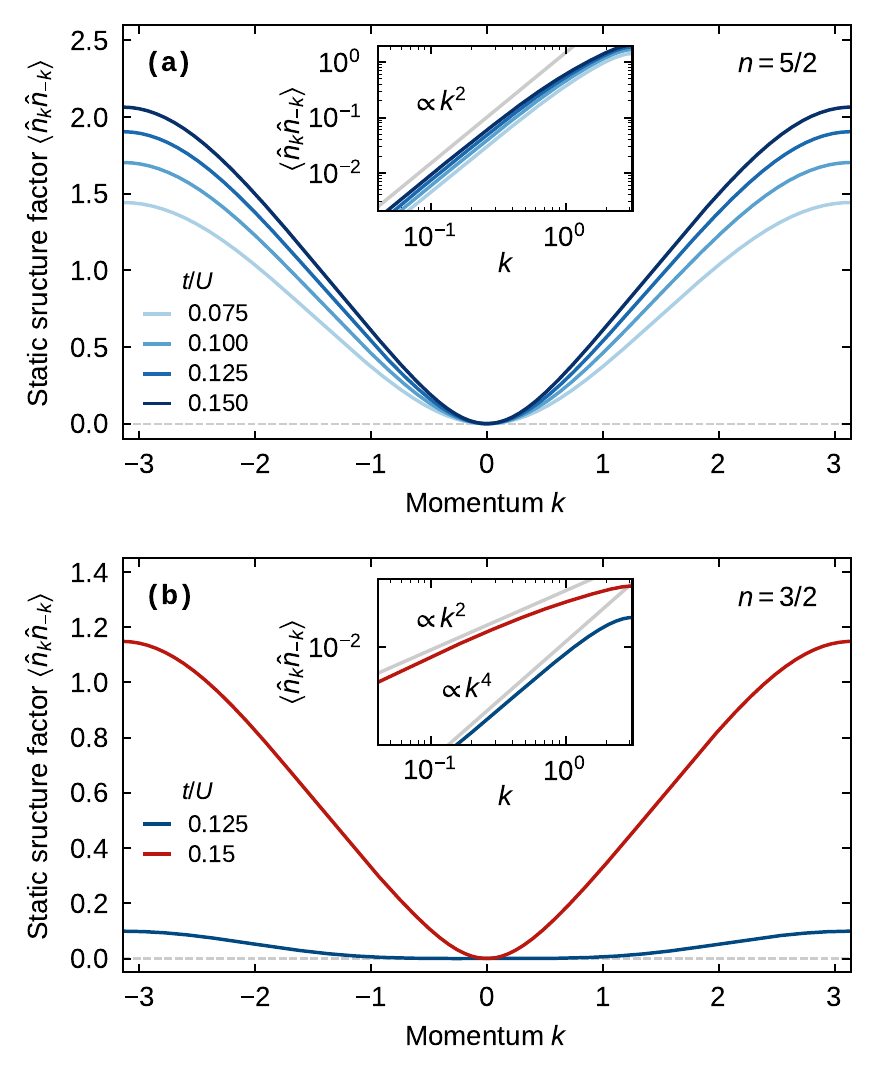}
    \caption{\label{fig:incommCompressibility}
        \textbf{Compressibility at non-integer filling.}
        The scaling of the static structure factor indicates finite compressibility at non-integer filling.
        (a)~Static structure factors at filling $n=5/2$.
        At sufficiently large $t/U$ the structure factor scales $\propto k^2$ for small momenta, compatible with the predictions from the quantum Lifshitz theory describing a compressible state.
        (b)~Static structure factors at filling for $n=3/2$.
        For small $t/U$ a transition to a Mott insulating state occurs, where we find a $\propto k^4$ scaling.
    }
\end{figure}

\textit{\textbf{Charge compressibility.---}}
To further substantiate the evidence for a compressible state on the numerically accessible scales, in the following, we consider the compressibility as determined by static density correlations. To this end, we again return to resolving charge- and dipole-conservation laws within our iDMRG approach. 

Our goal is to first understand what to expect of a state described by the Lifshitz model. In particular, the compressibility of the Lifshitz model in the absence of cosine terms is finite,
\begin{equation} \label{eq:3.2}
\begin{split}
    \kappa(k) &= C_{nn}(\omega=0,k) = \frac{1}{\pi^2}k^2 \braket{\phi(\omega=0,k)\phi(\omega=0,-k)} \\
    &= \frac{K}{\pi v}.
\end{split}
\end{equation}
As previously done for the dipole Luttinger liquid, we can further compute the associated equal-time density correlations. For the Lifshitz model, these are related to their static zero-frequency counterpart via
\begin{equation} \label{eq:3.3}
C_{nn}(\tau=0,k) = \frac{v}{2}k^2 C_{nn}(\omega=0,k) = \frac{K}{2\pi}k^2.
\end{equation}
The equal-time correlations of \eq{eq:3.3} can be determined efficiently in DMRG, and we should expect a $\sim k^2$ onset at small momenta when the state is described by a Lifshitz model.
In \fig{fig:incommCompressibility} we present $C_{nn}(\tau=0,k)$ as obtained numerically at the half-integer fillings $n=3/2$ and $n=5/2$. At sufficiently large hopping $t$, we indeed observe the quadratic onset $\sim k^2$ for small momenta.  This in turn is consistent with a constant $\lim_{k\to 0}C_{nn}(\omega=0,k)$ and thus a finite compressibility, as expected in the quantum Lifshitz model. We emphasize that independently of specific model assumptions, the observed $\sim k^2$ onset is markedly different from the $\sim |k|^3$ onset, that we have previously observed in the dipole Luttinger liquid at integer filling (\textit{cf.} \fig{fig:3}). Hence the density correlations  indicate a different ground state.

At the filling $n=3/2$ we additionally find an apparent Mott state with onset $\sim k^4$ of $C_{nn}(\tau=0,k)$ and exponentially decaying dipole correlations provided the hopping $t$ is sufficiently small. For the $n=5/2$ state, we also find such a Mott state, but it is located at very small $t$. We estimate the critical point of this transition for $n=3/2$ at $t/U \approx 0.14$ and for $n=5/2$ at $t/U\approx 0.02$. It would be interesting in the future to map out the transition between these two phases and determine whether an intermediate dipole Luttinger liquid exists at this filling. 

\begin{figure}
    \includegraphics[width=\columnwidth]{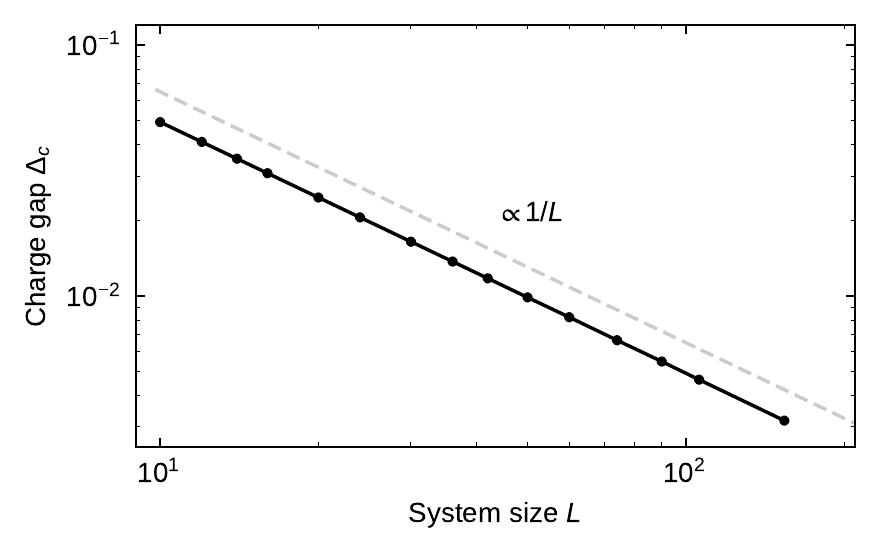}
    \caption{\label{fig:incommChargeGap}
        \textbf{Charge gap at non-integer filling.}
        Finite size flow of the charge excitation gap at filling $n=5/2$ and dipole hopping $t/U=0.125$, in good agreement with $\propto 1/L$ up to the accessible unit-cell sizes.
        A vanishing charge gap in the thermodynamic limit indicates a finite compressibility.
    }
\end{figure}

\textit{\textbf{Charge gap.---}}
Both the grand-canonical ground-state search and the static density correlations provide compelling evidence of the existence of a compressible state at non-integer filling at sufficiently large hopping $t$. As a final check, we investigate the energy gap $\Delta_c$ of charged single particle excitations as defined in \eq{eq:2.16}. If and only if the ground state is compressible, the charge gap vanishes in the limit of large systems: $\Delta_c \xrightarrow{L \rightarrow \infty} 0$. Specifically, for any system of length $L$, we find a finite-size charge gap whose scaling upon $L \rightarrow \infty$ we wish to determine. \fig{fig:incommChargeGap} shows the scaling of $\Delta_c$ for increasing system  sizes $L$ at half-integer filling  $n=5/2$ and dipole hopping $t/U=0.125$. Within our accessible computational resources, the associated finite size charge gap appears to close as $\Delta_c(L) \sim 1/L$ 
for large systems. Again, this apparently vanishing charge gap provides an indication for the compressibility of the ground state.
We note that in contrast to the charge gap $\Delta_c\sim 1/L$, the dipole excitation gap $\Delta_d$ in the Lifshitz model with dynamical exponent $z=2$ is expected to close as $\Delta_d \sim 1/L^2$, see Ref.~\cite{seiberg2022_lifshitz}.
Numerically, we verified that it becomes very small. For all probed system sizes, we find $\Delta_d/U \lesssim 10^{-5}$, making it unfeasible to capture the exact finite-size flow within the numerical accuracy for accessible bond dimensions.

\begin{figure}
    \includegraphics[width=\columnwidth]{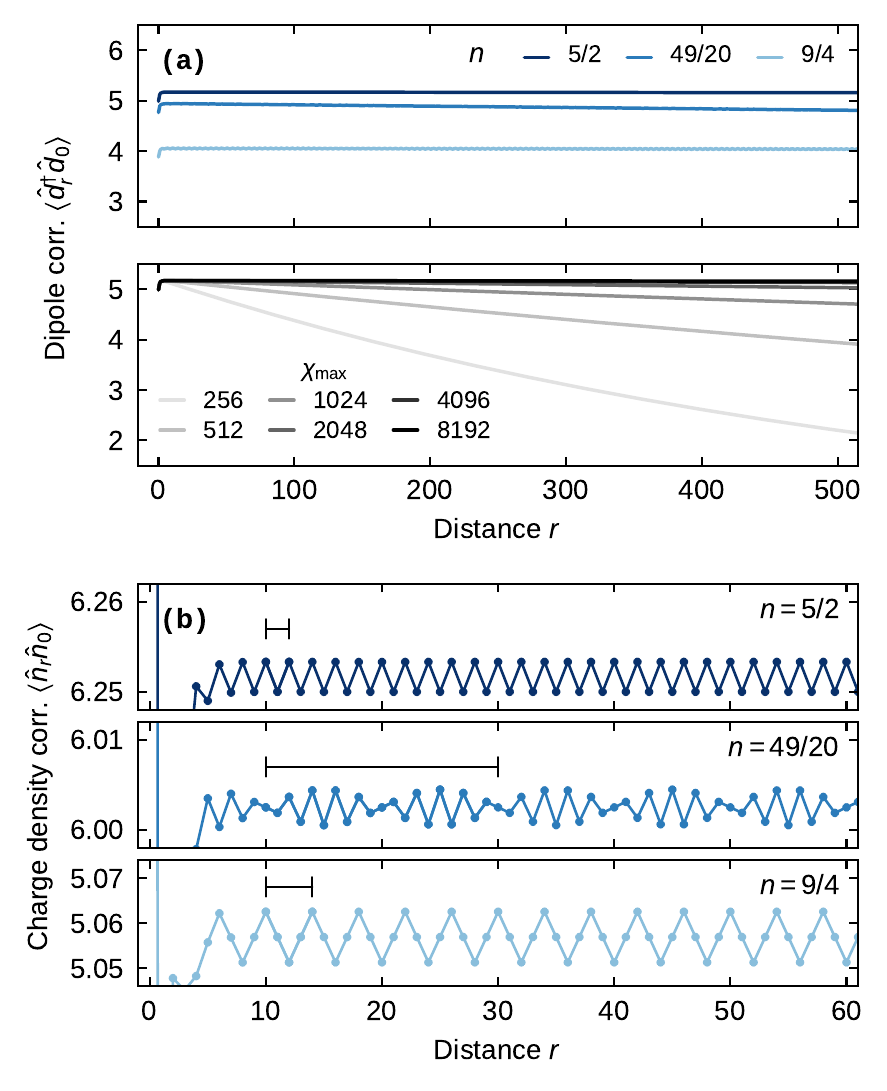}
    \caption{\label{fig:9}
        \textbf{Breaking of translation invariance and dipole long-range order at non-integer filling.}
        (a)~Upper panel: Dipole correlations $\langle \hat{d}^{\dagger}_{r}\hat{d}^{\phantom\dagger}_{0} \rangle$ at several non-integer rational fillings $n=p/q=5/2,\,49/20,\,9/4$, for $t/U=0.125$. The correlations remain constant even at large distances. 
        Lower panel: convergence in bond dimension for $n=5/2$.
        (b)~Charge density-density correlations $\langle \hat{n}_{r} \hat{n}_{0} \rangle$ exhibit a $q$ periodicity for the same fillings $n=p/q$ as in (a).
    }
\end{figure}

\textit{\textbf{Characterizing the compressible state: A dipole supersolid.---}}
The central property of the Lifshitz model \eq{eq:2.26} is the presence of off-diagonal long-range order in the dipole-dipole correlations functions. We recall that the dipole phase field $\theta_d(x)$ of the Luttinger liquid gets pinned to the gradient $-\partial_x \theta(x)$ of the charge phase field in the Lifshitz model, see \eq{eq:2.27}. The off-diagonal dipole correlations are long-ranged and are given by
\begin{equation}
    \braket{e^{i\partial_x \theta(x)}e^{-i\partial_x\theta(0)}} \xrightarrow{x\rightarrow \infty} e^{-\frac{\mathrm{const.}}{K}}.
\end{equation}
We verify this prediction numerically by computing $\braket{\hat{d}^\dagger_x\hat{d}_0}$ within iDMRG; \figc{fig:9}{a}. As the bond dimension is increased, the dipole correlations indeed approach a constant value on the accessible length scales of several  hundred sites. This indicates a spontaneous breaking of the dipole $U(1)$ symmetry, which is allowed even in one dimension due to a modified Mermin-Wagner theorem for systems with multipole conservation laws. Our numerical results show that the phenomenology of long-range dipole order is remarkably robust in the microscopic model \eq{eq:1.1}.
In addition, we verified numerically on finite system sizes that the dipole superfluid stiffness is finite (as is the case in the dipole Luttinger liquid). This can be probed by computing the sensitivity of the ground state energy to a twist in the boundary conditions~\cite{scalapino:1993}.

The presence of off-diagonal long-range order is not the only remarkable feature of our ground state at non-integer filling.
Quite generally, for a translation invariant system subject to both charge and dipole conservation, the ground state at filling $n=p/q \notin \mathds{N}$ and $p$, $q$ coprime is necessarily at least $q$-fold degenerate due to the non-commutativity of translations and dipole symmetry~\cite{seidel:2005}. The degenerate ground states are connected via translations. 
As a direct consequence, these states exhibit charge density wave (CDW) order with wave number $2\pi/q$. This feature is in agreement with the predictions of the quantum Lifshitz model of \eq{eq:rev2} in the absence of cosine terms. Since the correlator 
\begin{equation}
    \braket{e^{i\phi(r)}e^{-i\phi(0)}} \xrightarrow{r \rightarrow \infty} \mathrm{const.}
\end{equation}
exhibits long-range order, the density correlations feature long-range $q$ periodicity, cf. the expression \eq{eq:2.9} of the density in terms of the $\phi$-field in bosonization.
We thus expect to find density wave order for our system at any rational filling. \figc{fig:9}{b} demonstrates the presence of a CDW in the density-density correlations for different fillings $n=5/2,\,49/20,\,9/4$, confirming the expected associated periodicities $q=2,\,20,\,4$.
Interestingly, although the Lifshitz model describes a compressible state with vanishing charge gap, our mapping between boson and dipole occupation numbers introduced in Sec.~\ref{sec:I} remains valid. This is because the central criterion for its applicability, the bounded nature of charge fluctuations [see \eq{eq:2.3}] holds in the Lifshitz model of \eq{eq:rev2}. Formally, since $n(x) - n \sim \nabla \phi(x)$,
\begin{equation}
\begin{split}
    &\Braket{\Bigl(\int_0^x dx^\prime (n(x)-n)\Bigl)^2} = \\
    &= \braket{ (\phi(x)-\phi(0))^2 } \xrightarrow{x \rightarrow \infty} \mathrm{const.}\times K.
\end{split}
\end{equation}
Hence, cumulative charge fluctuations retain an area law. Since the dipole density remains well-defined, by virtue of $\partial_x q_d(x) = q(x)$ it inherits the DW order of the charge density. In quantum simulation platforms, both the bounded nature of charge fluctuations as well as the presence of DW order could be verified by sampling particle occupation number snapshots from the ground state wave function and using them to evaluate $\braket{n_d(x)n_d(0)}$. In line with this picture, we numerically observed small oscillations with period $q$ on top of the long-ranged dipole correlations $\braket{\hat{d}^\dagger_r\hat{d}_0}$ at non-integer rational fillings.

We conclude that a remarkable feature of this ground state at non-integer filling is the coexistence of a finite dipole superfluid stiffness and a
charge density wave order along with off-diagonal long-range order of the vortex operators that create local dipoles. As such, the state described by the Lifshitz model may be viewed as a \textit{dipole supersolid}.

\section{Conclusion and Outlook} \label{sec:outlook}
In this paper, we have investigated the ground-state quantum phases in the one-dimensional Bose-Hubbard model with dipole conservation. Utilizing the area-law nature of cumulative charge fluctuations in the ground states of the model, we were able to construct a local dipole density. This in turn, allowed us to develop an effective low-energy description of the system. At fixed integer boson densities, we found that the system undergoes a BKT transition between a gapped Mott state and a dipole Luttinger liquid that exhibits gapless particle-hole-type excitations, in agreement with iDMRG computations. The charge gap remains finite at integer filling with increasing hopping until an instability towards boson bunching is reached. At non-integer filling, however, our numerical results showed a ground state described by the one-dimensional quantum Lifshitz model, dubbed ``Bose-Einstein-insulator'' in Ref.~\cite{lake:2022}. This phase corresponds to a compressible state in which density wave order coexists with off-diagonal long-range order and finite superfluid stiffness for the dipole degrees of freedom. We therefore refer to this regime as a ``dipole supersolid''. General arguments suggest that this phase will eventually be unstable towards lattice effects. Nonetheless, the robustness of this compressible state within the unit-cell sizes accessible in our iDMRG approach is remarkable and suggests that the phenomenology of the dipole supersolid may be accessible in current quantum simulation platforms.

Collecting our results on the ground-state properties and charge/dipole energy gaps of the model \eq{eq:1.1} leads us to conclude with the $t/U-\mu/U$ -- phase diagram presented in \figc{fig:1}{b}: The system features Mott lobes with finite charge gap and integer boson filling, within which a transition between a fully gapped state and a dipole Luttinger liquid occurs. The lobes do not close until the bunching instability is reached. A special case is the $n=1$ Mott lobe, which remains in a fully gapped state up until bunching, which is why we focused mostly on $n=2$ in this paper.
The phase diagram shown in \fig{fig:1} is inferred from a parameter scan of $t/U$ and $\mu$ at for systems of 100 sites in the grand-canonical ensemble.

Open questions concerning the phase diagram of \fig{fig:1} -- beyond the eventual stability of the supersolid state -- exist in the regime of non-integer boson densities at small correlated hopping. There, we observed signatures of a transition between fractional filling Mott states and the compressible dipole supersolid (hatched areas). Mapping out the details of this potential transition is an interesting task for future work.

In addition, our results pave the way -- both analytically and numerically -- for tackling a number of related systems such as fermions or spin chains with dipole conservation. Specifically, our mapping to a model of microscopic dipoles may provide a good conceptual starting point for addressing questions about the non-equilibrium dynamics of excitations on top of the ground states obtained here. In future work, we plan to address such dynamical questions, including the evaluation of dynamic spectral functions. An additional question for future study is the fate of the microscopic dipole mapping at nonzero temperatures, where thermal fluctuations lead to a violation of the area law condition for charge fluctuations.  

Our results furthermore provide useful indications for potential experimental realizations of dipole phases beyond the simplest gapped Mott state. In particular, cold atoms in optical lattices in the presence of a strong linear tilt give rise to effective dipole-conserving dynamics and have been realized in Fermi-Hubbard systems both in one and two dimensions~\cite{guardado-sanchez:2020,scherg:2021,kohlert:2021,zahn:2022}. However, the associated correlated hopping strength $t/U=(t_{sp}/V)^2$ is generally suppressed by the ratio of the bare single-particle hopping strength $t_{sp}$ and the strength $V$ of the linear tilt~\cite{zhang_2020,scherg:2021}. Nonetheless, our analysis suggests that dipole Luttinger liquids or supersolids may already be accessible at moderate values of the correlated hopping. A more detailed investigation of the ground states and their transitions particularly at small $t$ is needed in order to substantiate this picture.

We emphasize further that through the mapping to microscopic dipoles, our work suggests potentially useful observables that can be used to analyze constrained models in experiments. In particular, quantum simulation platforms such as quantum gas microscopes have access to snapshots of the full system. This allows one to i) verify the conservation of the global dipole moment, ii) verify the area-law nature of cumulative charge fluctuations that guarantees a consistent local dipole density, and iii) perform the mapping to dipole degrees of freedom on the snapshots in order to study the dynamics of dipoles directly.

Beyond many-body systems in the presence of a linear tilt, dipole-conserving Hamiltonians similar to \eq{eq:1.1} are relevant to fractional quantum Hall systems placed on a thin cylinder~\cite{haldane1985_fqhe,trugman1985_fqh,berhgholtz2005_fqhe,seidel:2005,Sanjay19}. It would be very interesting to investigate whether the physics studied in our work can be of direct relevance to such setups.

\textit{Note added.} Reference~\cite{lake2022_dc} also provides an investigation of the phase diagram in the constrained Bose-Hubbard model.

\medskip
Data analysis and simulation codes are available on Zenodo upon reasonable request~\cite{zenodo}.

\begin{acknowledgments}
We thank Samuel Garratt, Johannes Hauschild, Clemens Kuhlenkamp, Leo Radzihovsky, Pablo Sala, and Zack Weinstein for many insightful discussions. We especially thank Jung Hoon Han, Byungmin Kang, and Ethan Lake for discussions about the static structure factor.
We acknowledge support from the Deutsche Forschungsgemeinschaft (DFG, German Research Foundation) under Germany’s Excellence Strategy--EXC--2111--390814868 and DFG Grants No. KN1254/1-2, KN1254/2-1, the European Research Council (ERC) under the European Union’s Horizon 2020 research and innovation programme (Grant Agreement No. 851161), as well as the Munich Quantum Valley, which is supported by the Bavarian state government with funds from the Hightech Agenda Bayern Plus. EA is supported by the NSF QLCI program through Grant No. OMA-2016245.
Matrix product state simulations were performed using the TeNPy package~\cite{hauschild:2018}.
\end{acknowledgments}

\appendix

\section{Computational methods} \label{app:A}
Throughout this paper, we employ tensor network methods to numerically access properties of the many-body ground state. 
We use the DMRG algorithm based on an MPS representation of the wave function, which allows for a controlled expansion in terms of the entanglement encoded in this ansatz~\cite{MurgReview,schollwock:2011}.
Starting from an initial product state, DMRG variationally optimizes the energy by local updates, where only $\chi_\mathrm{max}$ most important Schmidt states are kept.
After reaching convergence, we compute expectation values and correlation functions on the ground state MPS by contracting the relevant tensor networks.
For finite MPS we find that, in particular for large $t/U$, boundary effects become relevant to a point where they cannot be neglected.
Therefore, all our simulations utilize the infinite version of DMRG, directly working in the thermodynamic limit~\cite{vidal:2007}.

\subsection{Grand-canonical simulations}
In iDMRG, by fixing the size of the unit cell, one always implicitly imposes translation invariance.
Hence, it is typically very important that the periodicity of the ground state is commensurate with the unit cell.
For our dipole conserving model \eq{eq:1.1}, we are in a special situation when we work in the grand-canonical ensemble where the total particle number can fluctuate.
For any given unit-cell size $L$, the ansatz states are automatically at least $L$ periodic; however, because of the $q$-fold ground-state degeneracy, the ground state has a $2\pi/q$ periodicity at filling $p/q$.
Therefore the lowest-energy state is forced to have rational filling $p/L$, where $p$ and $L$ do not have to be coprime.
As a result, this leads to locking of the average density to rational fillings fractions, which appear as a staircase structure in grand-canonical cuts.
We verified that upon increasing $L$, the size and height of these plateaus reduce, signaling a tendency toward an incompressible state.

\subsection{Conservation laws}
In order to reach high bond dimensions and to be able to resolve dipole and charge sectors, we exploit the conservation of total charge $N$ and dipole moment $P$.
The construction of tensor networks symmetric under global transformations has been thoroughly discussed in the literature~\cite{singh:2010,singh:2011}.
Here, we briefly sketch the idea of how conservation laws are implemented, to then discuss how dipole conservation can be applied to infinite MPS algorithms.
In the context of fractional quantum Hall physics on thin cylinders, the momentum around the cylinder maps to the dipole moment of the particle density, and momentum conservation has been successfully exploited in this case~\cite{zaletel:2013}.

U(1) symmetries can be directly implemented on the level of the tensors $A_{\alpha\beta}^{[n], j_n}$ of the MPS representation
\begin{equation}
    \ket{\psi} = \sum_{\{j_{n}\}} \bigl[ \cdots A^{[n] j_{n}} A^{[n+1] j_{n+1}} \cdots \bigr] \ket{\dots, j_{n}, j_{n+1}, \dots},
\end{equation}
where for a unit cell of size $L$ we have $A^{[n]} = A^{[n+L]}$.
This is achieved by assigning quantum numbers, or charges, to the legs of the tensors.
Charges of contracted legs, i.e., the bonds in the MPS, are required to match, and tensors can carry charge themselves.
Then, MPS tensors $A^{[n], j_n}_{\alpha\beta}$ are constrained by the charge rule
\begin{equation}\label{eq:charge_rule}
    q^{[n]}_{\alpha} - q^{[n]}_{\beta} - q^{[n]}_{j_{n}} = Q^{[n]},
\end{equation}
where $q^{[n]}_{\alpha}$ ($q^{[n]}_{\beta}$) are the charges on the left (right) virtual leg, $q^{[n]}_{j_{n}}$ the charges of the physical leg, and $Q^{[n]}$ the total charge.
Only entries of the tensor for which the legs fulfill the charge rule can be nonzero.
The charge rule directly generalizes to tensors with any number of legs, such as the matrix product operators used to represent the Hamiltonian.
Generally, some convention for in- and out-going legs must be defined, specifying which legs connect to bra or ket states, which fixes the signs in \eq{eq:charge_rule}.
All tensor operations (e.g., permutation, reshaping, contractions, and decompositions) can be implemented to conserve the block structure imposed by the charge rules.
This can dramatically reduce the computational cost of tensor network algorithms, which in turn allows one to consider higher bond dimensions.

For our case of particle number and dipole conservation, we assign two sets of charges $(q^{[n]}_{N}, q^{[n]}_{P})$ to the tensor's legs, and nonzero elements of any tensor are only allowed for indices satisfying a charge rule for each set.
However, due to the fact that translations do not commute with the dipole operator, translating the MPS by $r$ sites does act non-trivially on the charges
\begin{equation}\label{eq:shift_rule}
    (q^{[n]}_{N}, q^{[n]}_{P}) \rightarrow (q^{[n]}_{N}, q^{[n]}_{P} + r * q^{[n]}_{N}).
\end{equation}
For operations within one unit cell this is not an issue, and all operations on tensors can be carried out as usual.
However, for operations on different unit cells, e.g., when optimizing the first or last tensor in iDMRG or for computing correlation functions, we have to make sure to apply the shift rule~\eq{eq:shift_rule} accordingly, and adjust the charges at every leg of the tensor.

With this modification iDMRG can immediately be applied to exploit dipole conservation.
Additionally, we can with that fix the dipole moment sector.
One important caveat to take into consideration is the ergodicity of iDMRG updates.
Due to the additional constraint arising from imposing dipole conservation, the variational space for optimizing the MPS ansatz is severely restricted, and fragments into sectors disconnected under standard two-site iDMRG updates~\cite{zaletel:2013}.
Hence, depending on the initial state the optimization may get stuck in a local minimum.
We mitigate this problem by using a subspace expansion method~\cite{hubig:2015} in combination with two-site iDMRG updates.
This introduces perturbations in the state and adds fluctuations to the quantum numbers, which significantly improves the ergodicity of iDMRG.


\begin{thebibliography}{71}
\bibitem [{foo()}]{footnote}%
  \BibitemOpen
  \href@noop {} {\bibinfo {title} {The dipole velocity $u_d$ is determined from
  a delicate finite-size flow of the dipole energy gap, while $k_d$ can be
  robustly determined from the decay of correlation functions.}}\BibitemShut
  {Stop}%
\end{thebibliography}
\end{document}